\documentclass{article}
\usepackage{bookstyle,bm,cite}
\begin{document}
\title{Cosmic Microwave Background Anisotropies up to Second Order}
%\runtitle{CMB Anisotropies up to Second Order}
\author{N. Bartolo$^1$, S. Matarrese$^1$ \& A. Riotto$^2$}
\address{$^1$ Dipartimento di Fisica ``G.\ Galilei'', 
        Universit\`{a} di Padova, 
        and INFN, Sezione di Padova, \\
        via Marzolo 8, I-35131 Padova, Italy \\
        $^2$ CERN, Theory Division, CH-1211 Geneva 23, Switzerland}

\frontmatter
\maketitle
\mainmatter%
%s1 ###

%\usepackage{latexsym}
%\usepackage[dvips]{graphicx} 

% PRIMORDIAL POTENTIAL
\newcommand{\fix}{\Phi(\mathbf{x})}
\newcommand{\fiLx}{\Phi_{\rm L}(\mathbf{x})}
\newcommand{\fiNLx}{\Phi_{\rm NL}(\mathbf{x})}
\newcommand{\fik}{\Phi(\mathbf{k})}
\newcommand{\fiLk}{\Phi_{\rm L}(\mathbf{k})}
\newcommand{\fiLkone}{\Phi_{\rm L}(\mathbf{k_1})}
\newcommand{\fiLktwo}{\Phi_{\rm L}(\mathbf{k_2})}
\newcommand{\fiLkthree}{\Phi_{\rm L}(\mathbf{k_3})}
\newcommand{\fiLkfour}{\Phi_{\rm L}(\mathbf{k_4})}
\newcommand{\fiNLk}{\Phi_{\rm NL}(\mathbf{k})}
\newcommand{\fiNLkone}{\Phi_{\rm NL}(\mathbf{k_1})}
\newcommand{\fiNLktwo}{\Phi_{\rm NL}(\mathbf{k_2})}
\newcommand{\fiNLkthree}{\Phi_{\rm NL}(\mathbf{k_3})}

% CONVOLUTION RELATED QUANTITIES
\newcommand{\kernel}{f_{\rm NL} (\mathbf{k_1},\mathbf{k_2},\mathbf{k_3})}
\newcommand{\dirac}{\delta^{(3)}\,(\mathbf{k_1+k_2-k})}
\newcommand{\dirackonektwokthree}{\delta^{(3)}\,(\mathbf{k_1+k_2+k_3})}

% ABBREVIATIONS FOR THE MOST USED COMMANDS
\newcommand{\beq}{\begin{equation}}
\newcommand{\eeq}{\end{equation}}
\newcommand{\beqarr}{\begin{eqnarray}}
\newcommand{\eeqarr}{\end{eqnarray}}

% ANGULAR PARTS
\newcommand{\angk}{\hat{k}}
\newcommand{\angn}{\hat{n}}

% TRANSFER FUNCTIONS
\newcommand{\tfnow}{\Delta_\ell(k,\tau_0)}
\newcommand{\tf}{\Delta_\ell(k)}
\newcommand{\tfone}{\Delta_{\ell_1}(k_1)}
\newcommand{\tftwo}{\Delta_{\ell_2}(k_2)}
\newcommand{\tfthree}{\Delta_{\ell_3}(k_3)}
\newcommand{\tffour}{\Delta_{\ell_1^\prime}(k)}
\newcommand{\deltatilde}{\tilde{\Delta}_{\ell_3}(k_3)}

% MULTIPOLES
\newcommand{\alm}{a_{\ell m}}
\newcommand{\almL}{a_{\ell m}^{\rm L}}
\newcommand{\almNL}{a_{\ell m}^{\rm NL}}
\newcommand{\almone}{a_{\ell_1 m_1}}
\newcommand{\almLone}{a_{\ell_1 m_1}^{\rm L}}
\newcommand{\almNLone}{a_{\ell_1 m_1}^{\rm NL}}
\newcommand{\almtwo}{a_{\ell_2 m_2}}
\newcommand{\almLtwo}{a_{\ell_2 m_2}^{\rm L}}
\newcommand{\almNLtwo}{a_{\ell_2 m_2}^{\rm NL}}
\newcommand{\almthree}{a_{\ell_3 m_3}}
\newcommand{\almLthree}{a_{\ell_3 m_3}^{\rm L}}
\newcommand{\almNLthree}{a_{\ell_3 m_3}^{\rm NL}}

% SPHERICAL HARMONICS
\newcommand{\YLMstar}{Y_{L M}^*}
\newcommand{\Ylmstar}{Y_{\ell m}^*}
\newcommand{\Ylmstarone}{Y_{\ell_1 m_1}^*}
\newcommand{\Ylmstartwo}{Y_{\ell_2 m_2}^*}
\newcommand{\Ylmstarthree}{Y_{\ell_3 m_3}^*}
\newcommand{\Ylmstarfour}{Y_{\ell_1^\prime m_1^\prime}^*}
\newcommand{\Ylmstarfive}{Y_{\ell_2^\prime m_2^\prime}^*}
\newcommand{\Ylmstarsix}{Y_{\ell_3^\prime m_3^\prime}^*}

\newcommand{\YLM}{Y_{L M}}
\newcommand{\Ylm}{Y_{\ell m}}
\newcommand{\Ylmone}{Y_{\ell_1 m_1}}
\newcommand{\Ylmtwo}{Y_{\ell_2 m_2}}
\newcommand{\Ylmthree}{Y_{\ell_3 m_3}}
\newcommand{\Ylmfour}{Y_{\ell_1^\prime m_1^\prime}}
\newcommand{\Ylmfive}{Y_{\ell_2^\prime m_2^\prime}}
\newcommand{\Ylmsix}{Y_{\ell_3^\prime m_3^\prime}}

% BESSEL FUNCTIONS
\newcommand{\jl}{j_\ell(k r)}
\newcommand{\jlfourone}{j_{\ell_1^\prime}(k_1 r)}
\newcommand{\jlfivetwo}{j_{\ell_2^\prime}(k_2 r)}
\newcommand{\jlsixthree}{j_{\ell_3^\prime}(k_3 r)}
\newcommand{\jlsix}{j_{\ell_3^\prime}(k r)}
\newcommand{\jlthree}{j_{\ell_3}(k_3 r)}
\newcommand{\jlthreetau}{j_{\ell_3}(k r)}

% GAUNT INTEGRALS
\newcommand{\Gaunt}{\mathcal{G}_{\ell_1^\prime \, \ell_2^\prime \, 
\ell_3^\prime}^{m_1^\prime m_2^\prime m_3^\prime}}
\newcommand{\Gaunttwo}{\mathcal{G}_{\ell_1^\prime \, \ell_2^\prime \, 
\ell_3}^{m_1^\prime m_2^\prime m_3}}
\newcommand{\Gauntstardef}{\mathcal{H}_{\ell_1 \, \ell_2 \, \ell_3}^{m_1 m_2 m_3}}
\newcommand{\Gauntstarone}{\mathcal{G}_{\ell_1 \, L \,\, \ell_1^\prime}
^{-m_1 M m_1^\prime}}
\newcommand{\Gauntstartwo}{\mathcal{G}_{\ell_2^\prime \, \ell_2 \, L}
^{-m_2^\prime m_2 M}}

% INTEGRATION VARIABLES
\newcommand{\dangn}{d \angn}
\newcommand{\dangk}{d \angk}
\newcommand{\dangkone}{d \angk_1}
\newcommand{\dangktwo}{d \angk_2}
\newcommand{\dangkthree}{d \angk_3}
\newcommand{\dk}{d^3 k}
\newcommand{\dkone}{d^3 k_1}
\newcommand{\dktwo}{d^3 k_2}
\newcommand{\dkthree}{d^3 k_3}
\newcommand{\dkfour}{d^3 k_4}
\newcommand{\dallk}{\dkone \dktwo \dk}

% FOURIER TRANSFORM
\newcommand{\FT}{ \int  \! \frac{d^3k}{(2\pi)^3} 
e^{i\mathbf{k} \cdot \angn \tau_0}}
\newcommand{\planewave}{e^{i\mathbf{k \cdot x}}}
\newcommand{\dallkfourier}{\frac{\dkone}{(2\pi)^3}\frac{\dktwo}{(2\pi)^3}
\frac{\dkthree}{(2\pi)^3}}

%BISPECTRA
\newcommand{\Bis}{B_{\ell_1 \ell_2 \ell_3}^{m_1 m_2 m_3}}
\newcommand{\Avbis}{B_{\ell_1 \ell_2 \ell_3}}

%LINE OF SIGHT INTEGRATION
\newcommand{\los}{\mathcal{L}_{\ell_3 \ell_1 \ell_2}^{L \, 
\ell_1^\prime \ell_2^\prime}(r)}
\newcommand{\loszero}{\mathcal{L}_{\ell_3 \ell_1 \ell_2}^{0 \, 
\ell_1^\prime \ell_2^\prime}(r)}
\newcommand{\losone}{\mathcal{L}_{\ell_3 \ell_1 \ell_2}^{1 \, 
\ell_1^\prime \ell_2^\prime}(r)}
\newcommand{\lostwo}{\mathcal{L}_{\ell_3 \ell_1 \ell_2}^{2 \, 
\ell_1^\prime \ell_2^\prime}(r)}
\newcommand{\losfNL}{\mathcal{L}_{\ell_3 \ell_1 \ell_2}^{0 \, 
\ell_1 \ell_2}(r)}

%%%%%%%%%%%%%%%%%%%%%%%%%%%%%%%%%%%%%%%%%%%%%%%%%%%%%%
\def\d{d}
\def\C{{\rm CDM}}
\def\me{m_e}
\def\te{T_e}
\def\ti{\tau_{\rm initial}}
\def\tci#1{n_e(#1) \sigma_T a(#1)}
\def\tr{\tau_R}
\def\dtr{\delta\tau_R}
\def\dd{\tilde\Delta^{\rm Doppler}}
\def\dsw{\Delta^{\rm Sachs-Wolfe}}
\def\clsw{C_l^{\rm Sachs-Wolfe}}
\def\cldop{C_l^{\rm Doppler}}
\def\Dt{\tilde{\Delta}}
\def\mut{\mu}
\def\vt{\tilde v}
\def\hp{ {\bf \hat p}}
\def\sdv{S_{\delta v}}
\def\svv{S_{vv}}
\def\bvt{\tilde{\bv}}
\def\delt{\tilde{\delta_e}}
\def\cos{{\rm cos}}
\def\nn{\nonumber \\}
\def\bq{ {\bf q} }
\def\ba{ {\bf p} }
\def\bap{ {\bf p'} }
\def\bqp{ {\bf q'} }
\def\bp{ {\bf p} }
\def\bpp{ {\bf p'} }
\def\bk{ {\bf k} }
\def\bx{ {\bf x} }
\def\bv{ {\bf v} }
\def\qp{ p^{\mu}k_{\mu} }
\def\qpp { p^{\mu} k'_{\mu} }
\def\bgm{ {\bf \gamma} }
\def\bkp{ {\bf k'} }
\def\gq{ g(\bq)}
\def\gqp{ g(\bqp)}
\def\fp{ f(\bp)}
\def\h#1{ {\bf \hat #1}}
\def\fpp{ f(\bpp)}
\def\fz{f^{(0)}(p)}
\def\fpz{f^{(0)}(p')}
\def\f#1{f^{(#1)}(\bp)}
\def\fps#1{f^{(#1)}(\bpp)}
\def\dq{ {d^3\bq \over (2\pi)^32E(\bq)} }
\def\dqp{ {d^3\bqp \over (2\pi)^32E(\bqp)} }
\def\dpp{ {d^3\bpp \over (2\pi)^32E(\bpp)} }
\def\dtq{ {d^3\bq \over (2\pi)^3} }
\def\dtqp{ {d^3\bqp \over (2\pi)^3} }
\def\dtpp{ {d^3\bpp \over (2\pi)^3} }
\def\part#1;#2 {\partial#1 \over \partial#2}
\def\deriv#1;#2 {d#1 \over d#2}
\def\Done{\Delta^{(1)}}
\def\Dtwo{\tilde\Delta^{(2)}}
\def\fone{f^{(1)}}
\def\ftwo{f^{(2)}}
\def\tg{T_\gamma}
\def\delpp{\delta(p-p')}
\def\delb{\delta_B}
\def\tc{\tau_0}
\def\DD{\langle|\Delta(k,\mu,\tau_0)|^2\rangle}
\def\DDL{\langle|\Delta(k=l/\tc,\mu)|^2\rangle}
\def\bkpp{{\bf k''}}
\def\kmkp{|\bk-\bkp|}
\def\kmkpsq{k^2+k'^2-2kk'x}
\def\tt{ \left({\tau' \over \tau_c}\right)}
\def\kt{ k\mu \tau_c}

%
%
%
%\renewcommand{\topfraction}{0.99}
%\renewcommand{\bottomfraction}{0.99}
%@twocolumnfalse\endcsname

\section{Preamble}
These lecture notes present the computation of the full system of   
Boltzmann equations describing the evolution 
of the photon, baryon and cold dark matter fluids up to second order
in perturbation theory, as recently studied in Refs.~\cite{paperI,paperII}. 
These equations allow to follow the time evolution of the cosmic microwave
background anisotropies at all angular scales
from the early epoch, when the cosmological perturbations were generated,
to the present, through the recombination era. The inclusion of second-order 
contributions is mandatory when one is interested in studying possible 
deviations from Gaussianity of cosmological perturbations, either of
primordial (e.g. inflationary) origin or due to their subsequent evolution. 
Most of the emphasis in these lectures notes will be given to the derivation
of the relevant equations for the study of cosmic microwave background
anisotropies and to their 
analytical solutions. 

%%%%%%%%%%%%%%%%%%%%%%%%%%%%%%%%%%%%%%%%%%%%%%%%%%%%%%%%%%%%%%%%%%%%%%%%%%%%
\section{Introduction}
%%%%%%%%%%%%%%%%%%%%%%%%%%%%%%%%%%%%%%%%%%%%%%%%%%%%%%%%%%%%%%%%%%%%%%%%%%%%

Cosmic Microwave Background (CMB) anisotropies play a special role 
in cosmology, as they allow an accurate 
determination of cosmological parameters and may provide 
a unique probe of the physics of the early universe and in particular of
the processes that gave origin to the primordial 
perturbations. 

Cosmological inflation \cite{lrreview} is nowadays considered 
the dominant paradigm for the generation of the initial seeds for 
structure formation. 
In the inflationary picture, 
the primordial cosmological perturbations are created from quantum 
fluctuations ``redshifted'' out of the horizon during an early period of 
accelerated expansion of the universe, where they remain ``frozen''.  
They are observable through CMB temperature anisotropies (and polarization) and
the large-scale clustering properties of the matter 
distribution in the Universe. 

This picture has recently received further spectacular confirmations 
from the  results of the Wilkinson Microwave Anisotropy Probe (WMAP) 
three year set of data \cite{wmap3}.
Since the observed cosmological perturbations are of the order
of $10^{-5}$, one might think that first-order perturbation theory
will be adequate for all comparisons with observations. This might not be
the case, though.  Present \cite{wmap3} and future experiments \cite{planck}  
may be sensitive to the non-linearities of the cosmological
perturbations at the level of second- or higher-order perturbation theory.
The detection of these non-linearities through the non-Gaussianity
(NG) in the CMB \cite{review} has become one of the
primary experimental targets. 

There is one fundamental  reason why a positive detection of NG 
is so relevant: it might help in discriminating among the various mechanisms
for the generation of the cosmological perturbations. Indeed,
various models of inflation, firmly rooted in modern 
particle physics theory,   predict a significant amount of primordial
NG generated either during or immediately after inflation when the
comoving curvature perturbation becomes constant on super-horizon scales
\cite{review}. While single-field  \cite{noi}
and two(multi)-field \cite{two} models of inflation predict
a tiny level of NG, ``curvaton''-type models,  in which
a  significant contribution to the curvature perturbation is generated after
the end of slow-roll inflation by the perturbation in a field which has
a negligible effect on inflation, may predict a high level of NG \cite{luw}.
Alternatives to the curvaton model are models 
where a curvature perturbation mode is  
generated by an inhomogeneity in the decay rate \cite{hk,varcoupling},
the mass  \cite{varmass} or the interaction rate \cite{thermalization}
of the particles responsible for the reheating after inflation. 
Other opportunities for generating the curvature perturbations occur
at the end of inflation \cite{endinflation}, during
preheating \cite{preheating}, and 
at a phase-transition producing cosmic strings \cite{matsuda}.

Statistics like the bispectrum and the trispectrum of the 
CMB can then be used to assess the level of primordial NG on 
various cosmological scales 
and to discriminate it from the one induced by 
secondary anisotropies and systematic effects \cite{review,hu,dt,jul}. 
A positive detection of a primordial NG in the CMB at some level 
might therefore 
confirm and/or rule out a whole class of mechanisms by which the cosmological
perturbations have been generated.

Despite the importance of NG in CMB anisotropies, little effort
has ben made so far to provide accurate
theoretical predictions of it. On the contrary, the vast majority of the
literature has been devoted to the computation of the bispectrum
of either the comovig curvature perturbation or the gravitational 
potential on large scales within given inflationary models. 
These, however, are not the  physical
quantities which are observed. One should instead provide a full prediction
for the second-order radiation transfer function. 
A preliminary step towards this goal has been taken in Ref.~\cite{fulllarge}
(see also \cite{twoT})
where the full second-order radiation transfer function 
for the CMB  anisotropies on large angular scales in a flat universe 
filled with matter and cosmological constant was computed, including 
the second-order generalization of the Sachs-Wolfe effect,
both the early and late Integrated Sachs-Wolfe (ISW) effects and the 
contribution of the second-order tensor modes.

There are many sources of NG in CMB anisotropies, beyond the primordial
one.  The most relevant  sources are the so-called  secondary anisotropies,
which arise  after the last scattering epoch. These anisotropies can be 
divided into two categories: scattering secondaries, when the CMB photons 
scatter with electrons along the line of sight, 
and gravitational secondaries when effects are mediated by 
gravity \cite{Hu:2001bc}.  Among the  scattering secondaries 
we may list the thermal Sunyaev-Zeldovich effect, where 
hot electrons in clusters transfer energy to the CMB photons, 
the kinetic Sunyaev-Zeldovich effect produced by the bulk motion of the 
electrons in clusters, the Ostriker-Vishniac effect, produced by bulk 
motions modulated by linear density perturbations, and effects due to  
reionization processes. The scattering secondaries 
are most significant on small angular scales
as density inhomogeneities, bulk and thermal motions  grow and become 
sizeable on small length-scales when structure formation proceeds.

Gravitational secondaries arise from the change 
in energy of photons when the gravitational potential is time-dependent, the 
ISW effect, and  gravitational lensing.  At late times, when the 
Universe becomes dominated by the dark energy, 
the gravitational potential on linear scales starts to decay, causing 
the ISW effect mainly on large angular scales. Other secondaries 
that result from a 
time dependent potential are the Rees-Sciama effect, produced during
the matter-dominated epoch by the time 
evolution of the potential on non-linear scales. 

The fact that the potential never grows appreciably means that most 
second order effects created by gravitational secondaries are generically small
compared to those created by 
scattering ones. However, when a photon propagates from the 
last scattering to us, its path may be deflected because of 
the gravitational lensing. This effect 
does not create anisotropies, but only modifies existing ones. Since
photons with large wavenumbers $k$  are lensed over many regions ($\sim k/H$, 
where $H$ is the Hubble rate) along the
line of sight, the corresponding second-order effect may be sizeable.
The  
three-point function arising from the correlation of the gravitational lensing 
and ISW effects generated by  
the matter distribution along the line of sight 
\cite{Seljak:1998nu,Goldberg:xm} and the Sunyaev-Zeldovich effect \cite{sk}
are large and detectable by Planck~\cite{ks}.

Another relevant source of NG comes from the physics operating at
the recombination. A naive estimate would tell that these non-linearities
are tiny being suppressed by an extra power of the gravitational 
potential. However, the dynamics  at recombination is quite involved because
all the non-linearities in the evolution of the baryon-photon fluid at 
recombination and the ones coming from general relativity should be 
accounted for. This complicated dynamics might lead
to  unexpected suppressions or enhancements of the NG at recombination. 
A step towards the evaluation of the three-point correlation function
has been taken in Ref. \cite{rec} where some effects were taken into account 
in the in so-called  squeezed triangle limit, corresponding to the case 
when one wavenumber is much smaller than the other two and was outside 
the horizon at recombination. 

These notes, which are based on Refs.~\cite{paperI,paperII}, 
present the computation of the full system of Boltzmann 
equations, describing the evolution of the photon, 
baryon and Cold Dark Matter (CDM) 
fluids, at second order and neglecting polarization, 
These equations allow to follow the time evolution of the CMB anisotropies 
at second order on all angular scales
from the early epochs, when the cosmological perturbations were generated,
to the present  time, through the recombination era. These calculations 
set the stage for the computation of the full second-order
radiation transfer function at all scales and for a 
a generic set of initial conditions specifying the level of primordial 
non-Gaussianity. 
Of course on small angular scales, fully non-linear
calculations of specific effects like Sunyaev-Zel'dovich,
gravitational lensing, etc.  would provide a more
accurate estimate of the resulting CMB anisotropy, however,
as long as the leading contribution to 
second-order statistics like the bispectrum  is
concerned, second-order perturbation theory suffices.

These notes are organized as follows. In Section 3 we provide the 
second-order metric and corresponding Einstein equations. In Section 4
the left-hand-side of the Boltzmann equation 
for the photon distribution function 
is derived at second order. The collision term is computed in Section 5. 
In Section 6 we present the second-order
Boltzmann equation for the photon brightness function, its formal
solution with the method of the integration along the line of sight and the
corresponding hierarchy equations for the multipole moments.
Section 7 contains the derivation of the Boltzmann equations at second order
for baryons and CDM. Section 8 contains the 
approximate solution of the Boltzmann equations up to first order. 
Section 9 contains a brief summary of the results. 
In Appendix A we give the explicit form of
Einstein's equations up to second order, while in Appendix B we 
provide the first-order solutions of Einstein's equations in various
cosmological eras. 

In performing the computation presented in these lecture notes, we have 
mainly followed Ref.~\cite{Dodelsonbook} (in particular chapter 4), 
where an excellent derivation of the Boltzmann equations for the 
baryon-photon fluid at first order is given, and Refs.~\cite{paperI,paperII}
for their second-order extension. 
Since the derivation at second order  is 
straightforward, but lenghty, the reader might benefit 
from reading the appropriate sections of Ref.~\cite{Dodelsonbook}. 
In the Conclusions (Section 9) we have also 
provided a Table which summarizes the many symbols 
appearing throughout these notes.

%%%%%%%%%%%%%%%%%%%%%%%%%%%%%%%%%%%%%%%%%%%%%%%%%%%%%%%%%%%%%%%%%%%%%%%%%%%%
\section{Perturbing gravity}
%%%%%%%%%%%%%%%%%%%%%%%%%%%%%%%%%%%%%%%%%%%%%%%%%%%%%%%%%%%%%%%%%%%%%%%%%%%%
 
Before tackling the problem of interest -- the computation of the Boltzmann
equations for the baryon-photon and CDM fluids -- we first provide the
necessary tools to deal with perturbed gravity, giving 
 the expressions for the Einstein tensor 
perturbed up to second order around  a flat 
Friedmann-Robertson-Walker background, and the relevant Einstein equations. 
In the following we will adopt the Poisson gauge which eliminates 
one scalar degree of freedom from the $g_{0i}$ component of the 
metric and one scalar and two vector degrees of freedom from $g_{ij}$. 
We will use a metric of the form   
\begin{equation}
\label{metric}
ds^2=a^2(\eta)\left[
-e^{2\Phi} d\eta^2+2\omega_i dx^i d\eta+(e^{-2\Psi}\delta_{ij}+\chi_{ij}) 
dx^i dx^j
\right]\, ,
\end{equation}
where $a(\eta)$ is the scale factor as a function of the conformal time 
$\eta$, and $\omega_i$ and $\chi_{ij}$ 
are the vector and tensor peturbation modes 
respectively. Each metric perturbation can be expanded into a 
linear (first-order) and a second-order part, as for example, the 
gravitational potential $\Phi=\Phi^{(1)}+\Phi^{(2)}/2$. However 
in the metric~(\ref{metric}) the choice of the exponentials greatly 
helps in computing the relevant expressions, and thus we will 
always keep them where it is convenient. From Eq.~(\ref{metric}) one 
recovers at linear order the well-known  
longitudinal gauge while at second order, one finds 
$\Phi^{(2)}=\phi^{(2)}-2 (\phi^{(1)})^2$ and $\Psi^{(2)}=
\psi^{(2)}+2(\psi^{(1)})^2$ where $\phi^{(1)}$, $\psi^{(1)}$ 
and $\phi^{(2)}$, $\psi^{(2)}$ (with 
$\phi^{(1)}=\Phi^{(1)}$ and $\psi^{(1)}=\Psi^{(1)}$) are the first and 
second-order gravitational 
potentials in the longitudinal (Poisson) gauge adopted in 
Refs.~\cite{MMB,review} as far as  scalar perturbations are concerned.
For the vector and tensor perturbations,  we will neglect linear 
vector modes since they are not produced in standard 
mechanisms for the generation of cosmological perturbations 
(as inflation), 
and we also neglect tensor modes at linear order, since they give a 
negligible contribution to second order 
perturbations. Therefore we take $\omega_i$ and $\chi_{ij}$ to be 
second-order vector and tensor perturbations of the metric. 

Let us now give our definitions for the connection coefficients and 
their expressions for the metric~(\ref{metric}). 
The number of spatial dimensions is $n=3$.
Greek indices ($\alpha, \beta, ..., \mu, \nu, ....$)
 run from 0 to 3, while latin indices ($a,b,\dots,i,j,k,\dots$,
$m,n,\dots$) run from 1 to 3. 
The space-time metric $g_{\mu \nu}$ has signature ($-,+,+,+$). 
The connection coefficients are defined as
\begin{equation}
\label{conness} \Gamma^\alpha_{\beta\gamma}\,=\,
\frac{1}{2}\,g^{\alpha\rho}\left( \frac{\partial
g_{\rho\gamma}}{\partial x^{\beta}} \,+\, \frac{\partial
g_{\beta\rho}}{\partial x^{\gamma}} \,-\, \frac{\partial
g_{\beta\gamma}}{\partial x^{\rho}}\right)\, .
\end{equation}
The Riemann tensor is defined as
\begin{equation}
R^{\alpha}_{~\beta\mu\nu}=
\Gamma^{\alpha}_{\beta\nu,\mu}-\Gamma^{\alpha}_{\beta\mu,\nu}+
\Gamma^{\alpha}_{\lambda\mu}\Gamma^{\lambda}_{\beta\nu}-
\Gamma^{\alpha}_{\lambda\nu}\Gamma^{\lambda}_{\beta\mu} \,.
\end{equation}
The Ricci tensor is a contraction of the Riemann tensor, 
$R_{\mu\nu}=R^{\alpha}_{~\mu\alpha\nu}$ 
and in terms of the connection coefficient it is given by
\begin{equation}
R_{\mu\nu}\,=\, \partial_\alpha\,\Gamma^\alpha_{\mu\nu} \,-\,
\partial_{\mu}\,\Gamma^\alpha_{\nu\alpha} \,+\,
\Gamma^\alpha_{\sigma\alpha}\,\Gamma^\sigma_{\mu\nu} \,-\,
\Gamma^\alpha_{\sigma\nu} \,\Gamma^\sigma_{\mu\alpha}\,.
\end{equation}
The Ricci scalar is the trace of the Ricci tensor, 
$R=R^{\mu}_{~\mu}$. The Einstein tensor is defined as
$G_{\mu\nu}=R_{\mu\nu}-\frac{1}{2}g_{\mu\nu}R$. 

The Einstein equations are written as 
$G_{\mu\nu}=\kappa^2 T_{\mu\nu}$, so that $\kappa^2=8\pi G_{\rm N}$, 
where $G_{\rm N}$ is the usual Newtonian gravitational constant.

%%%%%%%%%%%%%%%%%%%%%%%%%%%%%%%%%%%%%%%%%%%%%%%%%%%%%%%%%%%%%%%%
\section{The collisionless Boltzmann equation for photons}
%%%%%%%%%%%%%%%%%%%%%%%%%%%%%%%%%%%%%%%%%%%%%%%%%%%%%%%%%%%%%%%%
 
We are interested in the anisotropies in the cosmic
distribution of photons and inhomogeneities in the matter. Photons
are affected by gravity and by Compton scattering
with free electrons. The latter are tightly coupled to protons. Both
are, of course, affected by gravity. The metric which
determines the gravitational forces is influenced by all these components
plus CDM (and neutrinos). Our plan is to write down Boltzmann equations for 
the phase-space distributions of each species in the Universe.

The phase-space distribution of particles $g(x^i,P^\mu,\eta)$ 
is a function of spatial coordinates $x^i$, conformal time $\eta$, and 
momentum of the particle 
$P^\mu=dx^\mu/d\lambda$ where $\lambda$ parametrizes the particle path. 
Through the constraint $P^2 \equiv g_{\mu\nu} P^\mu P^\nu 
=- m^2$, where $m$ is 
the mass of the particle one can eliminate $P^0$ and $g(x^i,P^j,\eta)$ 
gives the number of particles in the differential phase-space volume 
$dx^1 dx^2 dx^3 dP^1 dP^2 dP^3$. 
In the following we will denote the 
distribution function for photons with $f$. 

The photons' distribution evolves according to the Boltzmann equation 
\begin{equation}
\label{Boltzgeneric}
\frac{df}{d\eta}= {\overline C}[f]\, ,
\end{equation}  
where the collision term is due to the scattering of photons off free 
electrons. In the following we will derive the left-hand side of 
Eq.~(\ref{Boltzgeneric}) while in the next section we will compute 
the collision term.

For photons we can impose $P^2 \equiv g_{\mu\nu} P^\mu P^\nu =0$ 
and using the metric~(\ref{metric}) in the conformal time $\eta$ we find
\begin{equation}
\label{P=0}
P^2=a^2\left[ - e^{2\Phi} (P^0)^2+\frac{p^2}{a^2} + 
2 \omega_i P^0 P^i\right]=0\, ,
\end{equation} 
where we define
\begin{equation}
\label{defp}
p^2= g_{ij}  P^iP^j\, .
\end{equation}
From the constraint~(\ref{P=0})
\begin{equation}
\label{P0}
P^0=e^{-\Phi}\left( \frac{p^2}{a^2}+2 \omega_i P^0P^i\right )^{1/2}\, .
\end{equation}
Notice that we immediately recover the usual zero and first-order relations
$P^0=p/a$ and $P^0=p(1-\Phi^{(1)})/a$. 

The components $P^i$ are proportional to $p n^i$, where $n^i$ is a unit 
vector with $n^in_i = \delta_{ij} n^in^j=1$.  
We can write $P^i=C n^i$, where $C$ is determined by
\begin{equation}
g_{ij} P^iP^j=C^2 \, a^2 (e^{-2\Psi}+\chi_{ij}n^in^j)=p^2\, ,
\end{equation}
so that 
\begin{equation}
\label{Pi}
P^i=\frac{p}{a} n^i\left(e^{-2\Psi}+\chi_{km}n^kn^m \right)^{-1/2} = 
\frac{p}{a} n^i e^{\Psi} \left( 1-\frac{1}{2} \chi_{km}n^kn^m \right)\, , 
\end{equation}
where the last equality holds up to second order in the perturbations. Again 
we recover the zero and first-order relations $P^i=p n^i/a$ and 
$P^i=p n^i (1+\Psi^{(1)})/a$ respectively. Thus up to second order we can write
\begin{equation}
\label{P0bis}
P^0=e^{-\Phi} \frac{p}{a} \left( 1+\omega_i\, n^i \right)\, .
\end{equation} 
Eq.~(\ref{Pi}) and~(\ref{P0bis}) allow us to replace $P^0$ and $P^i$ in terms 
of the variables $p$ and $n^i$. Therefore, as 
it is standard in the literature, from now on we will consider the phase-space 
distribution $f$ as a function of the momentum ${\bf p}=p n^i$ with magnitude 
$p$ and angular direction $n^i$, $f\equiv f(x^i, p, n^i,\eta)$. 

Thus, in terms of these variables, the total time derivative of the 
distribution function reads
\begin{equation}
\label{Df}
\frac{d f}{d \eta} = \frac{\partial f}{\partial \eta}+
\frac{\partial f}{\partial x^i} \frac{d x^i}{d \eta}+
\frac{\partial f}{\partial p} \frac{d p}{d \eta}+
\frac{\partial f}{\partial n^i} \frac{d n^i} {d \eta} \, . 
\end{equation}
In the following we will compute $d x^i/d \eta$, $d p/d \eta$ and 
$d n^i/d \eta$.
\\

\noindent
a) $d x^i/d \eta$:

From 
\begin{equation}
P^i=\frac{d x^i}{d \lambda}=\frac{d x^i}{d \eta} 
\frac{d\eta}{d\lambda}=\frac{d x^i}{d \eta} P^0
\end{equation}
and from Eq.~(\ref{Pi}) and~(\ref{P0bis}) 
\begin{equation}
\label{dxi}
\frac{d x^i}{d \eta}= n^i e^{\Phi+\Psi}
\left( 1- \omega_{j}\, n^j - \frac{1}{2} 
\chi_{km} n^k n^m \right)\, .
\end{equation}   

\noindent
b) $d p/d \eta$:

For $dp/d \eta$ we make use of the time component of the geodesic equation 
$d P^0/d \lambda= - \Gamma^0_{\alpha \beta} 
P^{\alpha} P^{\beta}$, where   
$d/d\lambda = (d\eta/d\lambda) 
\, d/d\eta=P^0 \, d/d\eta$, and  
\begin{equation}
\label{0geod}
\frac{d P^0}{ d \eta}= - \Gamma^0_{\alpha \beta} 
\frac{P^{\alpha} P^{\beta}}{P^0}\, ,
\end{equation}
Using the metric~(\ref{metric}) we find 
\begin{eqnarray}
\label{GPP}
2 \Gamma^0_{\alpha \beta} P^{\alpha} P^{\beta}&=& g^{0\nu} 
\left[ 2 \frac{\partial g_{\nu \alpha}}{\partial x^\beta} 
-\frac{\partial_{\alpha \beta}}{\partial x^\nu}
\right]P^\alpha P^\beta \nonumber \\
&=&2 ({\mathcal H}+\Phi')\left( P^0 \right)^2+4 \Phi_{,i} P^0P^i 
+4 {\mathcal H} \omega_i P^0 P^i \nonumber \\ 
&+&
2 e^{-2\Phi}\left[ ({\cal H}-\Psi') e^{-2 \Psi} \delta_{ij}  
-\omega_{i,j} + \frac{1}{2} \chi'_{ij}+{\cal H}\chi_{ij} \right] P^iP^j \, .
\nonumber \\
\end{eqnarray}
On the other hand the expression~(\ref{P0bis}) 
of $P^0$ in terms of $p$ and $n^i$ gives
\begin{eqnarray}
\frac{dP^0}{d \eta}&=&-\frac{p}{a} \frac{d \Phi}{d \eta} e^{-\Phi} 
\left( 1+ \omega_i n^i
\right)+e^{-\Phi} \left( 1+\omega_i \, n^i \right) \frac{d(p/a)}{d\eta}
\nonumber \\
&+&
\frac{p}{a} e^{-\Phi} \frac{d(\omega_i \, n^i)}{d\eta}\, .
\end{eqnarray} 
Thus Eq.~(\ref{0geod}) allows us express $dp/d\eta$ as
\begin{eqnarray}
\label{dp1}
\frac{1}{p} \frac{dp}{d\eta}=- {\cal H} +\Psi'-\Phi_{,i} \,n^i e^{\Phi+\Psi} 
- \omega'_{i}\,n^i-\frac{1}{2} \chi'_{ij}n^in^j\, ,
\end{eqnarray} 
where in Eq.~(\ref{GPP}) we have replaced $P^0$ and $P^i$ by 
Eqs.~(\ref{P0bis}) and 
(\ref{Pi}). Notice that in order to obtain Eq.~(\ref{dp1}) 
we have used the following 
expressions for the total time derivatives of the metric perturbations
\begin{eqnarray}
\label{dPhi}
\frac{d \Phi}{d\eta}&=&\frac{\partial \Phi}{\partial \eta}+
\frac{\partial \Phi}{\partial x^i} \frac{d x^i}{d \eta} 
\nonumber \\
&= &
\frac{\partial \Phi}{\partial \eta}+\frac{\partial \Phi}{\partial x^i}n^i 
e^{\Phi+\Psi}\left( 1-\omega_j\, n^j -\frac{1}{2} \chi_{km}n^k n^m \right)
\end{eqnarray}
and 
\begin{eqnarray}
\frac{d (\omega_{i} n^i)}{d \eta}=n^i\left( 
\frac{\partial \omega_i}{\partial \eta}+
\frac{\partial \omega_i}{\partial x^j} \frac{dx^j}{d\eta}\right)=
\frac{\partial \omega_i}{\partial \eta}n^i+
\frac{\partial \omega_i}{\partial x^j} n^i n^j \, ,
\end{eqnarray}
where we have taken into account that $\omega_i$ is already a second-order 
perturbation 
so that we can neglect $dn^i/d\eta$ which is at least a first order quantity, 
and we can 
take the zero-order expression in Eq.~(\ref{dxi}), $dx^i/d\eta=n^i$.     
In fact there is also an alternative expression for 
$dp/d\eta$ which turns out to be useful later and which can be obtained by 
applying once more Eq.~(\ref{dPhi}) 
\begin{eqnarray}
\label{dp2}
\frac{1}{p}\frac{dp}{d\eta}=-{\cal H} -\frac{d\Phi}{d\eta}
+\Phi'+\Psi'-\omega'_{i}\,n^i
-\frac{1}{2}\chi'_{ij} n^in^j\, .
\end{eqnarray}
\\ 
\\
c) $d n^i/d \eta$: 
\\
\\
We can proceed in a similar way to compute $dn^i/d\eta$. 
Notice that since in Eq.~(\ref{Df}) it multiplies 
$\partial f/\partial n^i$ which is first order, we need only 
the first order perturbation of $dn^i/d\eta$.  
We use the spatial components of the geodesic equations $dP^i/d\lambda=- 
\Gamma^i_{\alpha \beta}P^{\alpha}P^{\beta}$ written as
\begin{eqnarray}
\label{geoi}
\frac{dP^i}{d\eta}=-\Gamma^i_{\alpha \beta}\frac{P^{\alpha}P^{\beta}}{P^0}\, .
\end{eqnarray}
For the right-hand side we find, up to second order,
\begin{eqnarray}
\label{RHSgeoi}
&&2\Gamma^i_{\alpha \beta} P^{\alpha}P^{\beta}=g^{i\nu} \left[ 
\frac{\partial g_{\alpha\nu}}{\partial x^\beta} +
\frac{\partial g_{\beta \nu}}{\partial x^{\alpha}}
-\frac{\partial g_{\alpha \beta}}{\partial x^\nu}\right] 
P^\alpha P^\beta \\
&&= 4 ({\cal H}-\Psi')P^i P^0 +2\left( \chi^{i\prime}_{~k}+
\omega^i_{,k}-\omega_k^{~,i} \right) P^0P^k \nonumber \\
&&+ 
\left(2 \frac{\partial \Phi}{\partial x^i} e^{2\Phi+2\Psi} 
+2 \omega^{i\prime} +2 {\cal H} \omega^i \right)
\left( P^0 \right)^2 \nonumber - 4 \frac{\partial 
\Psi}{\partial x^k} P^i P^k \nonumber \\
&&+ 2 \frac{\partial \Psi}{\partial x^i}\delta_{km} P^kP^m
- \left[ 2{\cal H} \omega^i \delta_{jk}-
\left(\frac{\partial\chi^i_{~j}}{\partial x^k} +
\frac{\partial\chi^i_{~k}}{\partial x^j} - 
\frac{\partial\chi_{jk}}{\partial x_i }\right) \right]P^jP^k
\, \nonumber, 
\end{eqnarray}
while the expression~(\ref{Pi}) of $P^i$ in terms of our variables $p$ 
and $n^i$ in the left-hand side of Eq.~(\ref{geoi}) brings
\begin{eqnarray}
\frac{dP^i}{d\eta} &=&\frac{p}{a} e^{\Psi}\left[ \frac{d\Psi}{d\eta} n^i+
\frac{a}{p} \frac{d(p/a)}{d\eta} n^i+\frac{dn^i}{d\eta} \right]
\left(1-\frac{1}{2} \chi_{km} n^kn^m \right) \nonumber \\
&-&\frac{p}{a}n^i e^{\Psi} \frac{1}{2} 
\frac{d \left(\chi_{km}n^kn^m\right)}{d\eta} \, . 
\end{eqnarray}
Thus, using the expression~(\ref{Pi}) for $P^i$ and~(\ref{P0}) for $P^0$ in 
Eq.~(\ref{RHSgeoi}), together with the previous result~(\ref{dp1}), 
the geodesic equation~(\ref{geoi}) gives the following 
expression $dn^i/d\eta$ (valid up to first order)
\begin{eqnarray}
\label{dni}
\frac{d n^i}{d\eta}=\left( \Phi_{,k}+
\Psi_{,k} \right) n^k n^i-\Phi^{,i}-\Psi^{,i}\, .
\end{eqnarray}

To proceed further we now expand the distribution function for 
photons around the zero-order value $f^{(0)}$ which is that of a Bose-Einstein 
distribution
\begin{equation}
\label{BEd}
f^{(0)}(p,\eta)=2\,\, \frac{1}{\exp\left\{\frac{p}{T(\eta)}\right\}-1}\, ,
\end{equation}
where $T(\eta)$ is the average (zero-order) temperature and the factor $2$ 
comes from the spin degrees of photons. The perturbed 
distribution of photons will depend also on $x^i$ and on the propagation 
direction $n^i$ so as to account for inhomogeneities and anisotropies
\begin{equation}
\label{expf}
f(x^i,p,n^i,\eta)=f^{(0)}(p,\eta)+f^{(1)}(x^i,p,n^i,\eta)+\frac{1}{2} 
f^{(2)}(x^i,p,n^i,\eta)\, ,
\end{equation} 
where we split the perturbation of the distribution function into a 
first and a second-order part. The Boltzmann equation 
up to second order can be written in a straightforward way by 
recalling that the total time derivative of a given $i$-th perturbation, as 
{\it e.g.} $df^{(i)}/d\eta$ is {\it at least} a quantity of the $i$-th order. 
Thus it is easy to realize, looking at Eq.~(\ref{Df}), that 
 the left-hand side of Boltzmann equation can be written 
up to second order as 
\begin{eqnarray}
\label{LHSBoltz}
\frac{df}{d\eta}&=&
\frac{d f^{(1)}}{d\eta}+\frac{1}{2} \frac{df^{(2)}}{d\eta}
-p\frac{\partial f^{(0)}}{\partial p}
\frac{d}{d\eta}\left(\Phi^{(1)}+\frac{1}{2} 
\Phi^{(2)}\right) \nonumber \\
&+&
p \frac{\partial f^{(0)}}{\partial p} \frac{\partial }{\partial \eta}
\left( \Phi^{(1)}+\Psi^{(1)}+\frac{1}{2} \Phi^{(2)}+
\frac{1}{2} \Psi^{(2)} \right) \nonumber \\
&-& p \frac{\partial f^{(0)}}{\partial p} 
\frac{\partial \omega_i}{\partial \eta}n^i
- \frac{1}{2} p \frac{\partial f^{(0)}}{\partial p} \frac{\partial \chi_{ij}} 
{\partial \eta} n^in^j\, ,
\end{eqnarray}
where for simplicity in Eq.~(\ref{LHSBoltz}) we have already used the 
background Boltzmann equation $(df/d\eta)|^{(0)}=0$. 
In Eq.~(\ref{LHSBoltz}) there are all the terms which will give rise to 
the integrated Sachs-Wolfe 
effects (corresponding to the terms which explicitly 
depend on the gravitational perturbations), 
while other effects, such as the gravitational lensing, are still 
hidden in the (second-order part) of the first term. In fact in order to 
obtain Eq.~(\ref{LHSBoltz}) we just need for the time being  
to know the expression for $dp/d\eta$, Eq.~(\ref{dp2}). 
%%%%%%%%%%%%%%%%%%%%%%%%%%%%%%%%%%%%%%%%%%%%%%%%%%%%%%%%%%%%%%%%%%%%%%%
\section{Collision term}
%%%%%%%%%%%%%%%%%%%%%%%%%%%%%%%%%%%%%%%%%%%%%%%%%%%%%%%%%%%%%%%%%%%%%%%%
 
\subsection{The Collision Integral}
\label{CI}

In this section we focus on the collision term due to Compton scattering 
\begin{equation}
e({\bf q}) \gamma({\bf p}) \longleftrightarrow e({\bf q}') 
\gamma({\bf p}')\, ,
\end{equation}
where we have indicated the momentum of the photons  and electrons involved in 
the collisions. The collision term will be important for 
small scale anisotropies and 
spectral distortions. The important point to compute the 
collision term is that for 
the epoch of interest very little energy is transferred. Therefore one can 
proceed by expanding the right hand side of Eq.~(\ref{Boltzgeneric}) 
both in the small perturbation, Eq.~(\ref{expf}), and in the small energy 
transfer. Part of the computation up to second order has been done in 
Refs.~\cite{HSS,DJ,Huthesis} (see also \cite{roy}). In particular 
Refs.~\cite{HSS,DJ} are focused on the effects 
of reionization on the CMB anisotropies thus keeping in the collision 
term those contributions which are relevant for the small-scale 
effects due to reionization and neglecting 
the effects of the metric perturbations on the left-hand side of 
Eq.~(\ref{Boltzgeneric}). 
We will mainly follow the formalism of Ref.~\cite{DJ} and we will 
keep all the terms arising 
from the expansion of the collision term up to second order.   

The collision term is given (up to second order) by
\begin{eqnarray}
\label{collisionterm0}
{\overline C}({\bf p})=C({\bf p})a e^{\Phi}\, ,
\end{eqnarray}
where $a$ is the scale factor and
\footnote{The reason why we write the collision term as 
in Eq.~(\ref{collisionterm0}) is that the starting point of the 
Boltzmann equation requires differentiation with respect to an 
affine parameter $\lambda$, $df/d\lambda=C'$. 
In moving to the conformal time $\eta$ one rewrites the Boltzmann 
equation as $df/d\eta=C'(P^{0})^{-1}$, 
with $P^0=d\eta/d\lambda$ given by Eq.~(\ref{P0bis}). Taking 
into account that the collision term is at least of first order, 
Eq.~(\ref{collisionterm0}) then follows.}
\begin{eqnarray}
\label{collisionterm}
C({\bf p})&=&\frac{1}{E({\bf p})} \int \frac{d{\bf q}}{(2 \pi)^3 2E({\bf q})} 
\frac{d{\bf q}'}{(2 \pi)^3 2E({\bf q}')} 
\frac{d{\bf p}'}{(2 \pi)^3 2E({\bf p}')} \nonumber \\
&\times & (2\pi)^4 \delta^4(q+p-q'-p') \left| M \right|^2  \nonumber \\
& \times & \{ g({\bf q}')f({\bf p}')[ 1+f({\bf p})]-
g({\bf q})f({\bf p})[ 1+f({\bf p}')]\} 
\end{eqnarray}
where $E({\bf q})=(q^2+m_e^2)^{1/2}$, $M$ is the amplitude of 
the scattering process,
$\delta^4(q+p-q'-p')=\delta^3({\bf q}+{\bf p}-{\bf q}'-{\bf p}') 
\delta(E({\bf q})+p-E({\bf q}') -p')$ 
ensures the energy-momentum conservation and $g$ is 
the distribution function for electrons. The Pauli suppression factors $(1-g)$
have been dropped since for the epoch of interest the density of 
electrons $n_e$ is low. 
The electrons are kept in thermal equilibrium by Coulomb 
interactions with protons and 
they are non-relativistic, thus we can take a Maxwell-Boltzmann 
distribution around some bulk velocity ${\bf v}$
\begin{eqnarray}
\label{gel}
g({\bf q})=n_e \left( \frac{2 \pi}{m_e T_e}\right)^{3/2} 
\exp\left\{-\frac{({\bf q}-m_e{\bf v})^2}{2m_e T_e} \right\} 
\end{eqnarray}      
By using the three dimensional delta function the energy transfer is given by 
$E({\bf q})-E({\bf q}+{\bf p}-{\bf p}')$ and it turns out to be 
small compared to the typical thermal energies
\begin{equation}
\label{par}
E({\bf q})-E({\bf q}+{\bf p}-{\bf p}')\simeq \frac{({\bf p}-{\bf p}') 
\cdot {\bf q}}{m_e}
={\cal O}(Tq/m_e)\, . 
\end{equation}  
In Eq.~(\ref{par}) we have used $E({\bf q})=m_e+q^2/2m_e$ and the fact 
that, since the scattering 
is almost elastic ($p\simeq p'$), $({\bf p}-{\bf p'})$ is of order 
$p\sim T$, with $q$ much bigger 
than $({\bf p}-{\bf p'})$. In general, 
the electron momentum has two contributions, 
the bulk velocity ($q=m_e v$)  and the thermal 
motion ($q \sim (m_e T)^{1/2}$) and thus the parameter expansion 
$q/m_e$ includes 
the small bulk velocity ${\bf v}$ and the ratio 
$(T/m_e)^{1/2}$ which is small because 
the electrons are non-relativistic. 

The expansion of all the quantities entering the collision term 
in the energy transfer parameter and the integration over the momenta 
${\bf q}$ and 
${\bf q}'$ is described in details in Ref.~\cite{DJ}. 
It is easy to realize that we just need the scattering amplitude 
up to first order since at zero order 
$g({\bf q}')=g({\bf q}+{\bf p}-{\bf p}')=g({\bf q})$ 
and $\delta(E({\bf q})+p-E({\bf q}')-p')=\delta(p-p')$ so that 
all the zero-order quantities 
multiplying $\left| M \right|^2$ vanish. To first order 
\begin{equation}
\left| M \right|^2=6\pi\sigma_T 
m_e^2[(1+\cos^2\theta)-2\cos\theta(1-\cos\theta){\bf q}\cdot 
({\bf \hat{p}}+{\bf \hat{p}}')/m_e]\, ,
\end{equation}
where $\cos\theta={\bf n} \cdot {\bf n'}$ is the 
scattering angle and $\sigma_T$ the 
Thompson cross-section. The resulting collision term up to second order 
is given by~\cite{DJ} 
\begin{eqnarray}
\label{Integralcolli}
C(\bp) &=&  
{3n_e\sigma_T\over 4p} \int dp' p' {d\Omega' \over 4 \pi}
\bigg[ c^{(1)}(\bp,\bpp) + c^{(2)}_\Delta(\bp,\bpp)+c^{(2)}_v(\bp,\bpp) 
\nonumber \\
& +& c^{(2)}_{\Delta v}(\bp,\bpp) + 
c^{(2)}_{vv}(\bp,\bpp)+c^{(2)}_K(\bp,\bpp)   \bigg]\, ,
\end{eqnarray}
where we arrange the different contributions following Ref.~\cite{DJ}.
The first order term reads 
\begin{eqnarray}
c^{(1)}(\bp,\bpp) &=& (1+\cos^2\theta)\Bigg[
\delta(p-p') (\fps1-\f1) \nonumber \\
&+&(\fpz-\fz) (\bp-\bpp)\cdot\bv {\partial \delta(p-p')
			\over \partial p'} \Bigg]\, ,
\end{eqnarray}
while the second-order terms  have been separated into four parts. 
There is the so-called anisotropy suppression term 
\begin{eqnarray}
c^{(2)}_\Delta(\bp,\bpp) =\frac{1}{2} \left(1+\cos^2\theta\right)
		 \delta(p-p')(\fps2 - \f2)\, ;
\end{eqnarray}
a term which depends on the second-order velocity 
perturbation defined by the expansion 
of the bulk flow as ${\bf v}={\bf v}^{(1)}+{\bf v}^{(2)}/2$
\begin{equation}
c^{(2)}_v(\bp,\bpp)=\frac{1}{2}(1+\cos^2 \theta)  
(\fpz-\fz) (\bp-\bpp)\cdot\bv^{(2)}\, {\partial \delta(p-p')
			\over \partial p'}\, ;
\end{equation}
a set of terms coupling the photon perturbation to the velocity
\begin{eqnarray}
c^{(2)}_{\Delta v}(\bp,\bpp)&=&
\left(\fps1 - \f1\right)
\Bigg[  \left(1+\cos^2\theta\right)(\bp-\bpp)\cdot\bv \nonumber \\
&\times& {\partial \delta(p-p') \over \partial p'} 
- 2\cos\theta(1-\cos\theta) \delta(p-p')
		({\bf n}+ {\bf n}')\cdot \bv
\Bigg]\, , \nonumber
\end{eqnarray}
and a set of source terms quadratic in the velocity
\begin{eqnarray}
c^{(2)}_{vv}(\bp,\bpp) & = &
	\left(\fpz-\fz\right)\ (\bp-\bpp)\cdot\bv
\Bigg[ \left(1+\cos^2\theta\right) \nonumber \\
&\times & {(\bp-\bpp)\cdot\bv\over2}
		 {\partial^2 \delta(p-p')
			\over \partial p'^2} \nonumber \\
&-& 2\cos\theta(1-\cos\theta)
({\bf n}+ {\bf n}')\cdot \bv{\partial \delta(p-p')
			\over \partial p'} \Bigg]\,. \nonumber \\
\end{eqnarray}
The last contribution are the Kompaneets terms 
describing spectral distortions to the CMB
\begin{eqnarray}
c^{(2)}_K(\bp,\bpp)& = & \left(1+\cos^2\theta\right) {(\bp-\bpp)^2\over2\me}
	\Bigg[  \left( \fpz-\fz\right) \te \\
&\times & {\partial^2 \delta(p-p')
			\over \partial p'^2} 
- \left(\fpz+\fz+2\fpz\fz\right) \nonumber \\
& \times & {\partial \delta(p-p')
			\over \partial p'} \Bigg] 
+ {2(p-p')\cos\theta(1-\cos^2\theta)\over\me}
	\Bigg[ \delta(p-p') \nonumber \\
& \times & \fpz (1+\fz) \left(\fpz-\fz\right){\partial \delta(p-p')
			\over \partial p'} \Bigg]\, . \nonumber 
\end{eqnarray}
Let us make a couple of comments about the various contributions to the 
collision term. First, 
notice the term $c^{(2)}_v(\bp,\bpp)$ due to second-order 
perturbations in the velocity of electrons which is absent in Ref.~\cite{DJ}. 
In standard cosmological scenarios (like inflation) vector perturbations 
are not generated at 
linear order, so that linear velocities are irrotational $v^{(1)i}=\partial^i 
v^{(1)}$. However at 
second order vector perturbations are generated after horizon crossing as 
non-linear combinations 
of primordial scalar modes. Thus we must take into account
also a transverse (divergence-free) component, $v^{(2)i}=\partial^i v^{(2)}+ 
v^{(2)i}_T$ with 
$\partial_i v^{(2)i}_{T}=0$. As we will see such vector perturbations will 
break azimuthal symmetry 
of the collision term with respect to a given mode ${\bf k}$, which instead  
usually holds at linear order. Secondly, notice that the 
number density of electrons appearing in Eq.~(\ref{Integralcolli}) must be 
expanded as 
$n_e = \bar{n}_e(1+\delta_e)$ and then 
\begin{equation}
\label{deltac1}
\delta^{(1)}_e \,c^{(1)}(\bp,\bpp)
\end{equation}
gives rise to second-order contributions in addition to the list above, 
where we split $\delta_e=\delta^{(1)}_e+\delta^{(2)}_e/2$ into a 
first- and second-order part. In particular the 
combination with the term proportional to ${\bf v}$ in $c^{(1)}(\bp,\bpp)$ 
gives rise to the so-called Vishniac effect, as discussed in Ref.~\cite{DJ}.  

%%%%%%%%%%%%%%%%%%%%%%%%%%%%%%%%%%%%%%%%%%%%%%%%%%%%%%%%%%%%%%%%%%%%%%%%%%%
\subsection{Computation of different contributions to the collision term}
%%%%%%%%%%%%%%%%%%%%%%%%%%%%%%%%%%%%%%%%%%%%%%%%%%%%%%%%%%%%%%%%%%%%%%%%

In the integral~(\ref{Integralcolli}) over the momentum ${\bf p}'$ the 
first-order term gives the usual collision term
\begin{equation}
\label{C1}
C^{(1)}({\bf p})=n_e \sigma_T \left[ f^{(1)}_0(p)+\frac{1}{2}f^{(1)}_2 
P_2({\bf \hat v} \cdot 
{\bf n})-f^{(1)}-p\frac{\partial f^{(0)}}{\partial p} {\bf v} \cdot {\bf n} 
\right]\, ,
\end{equation}
where one uses the decomposition in Legendre polynomials
\begin{equation} 
\label{dec1}
f^{(1)}({\bf x},p,{\bf n})=\sum_\ell (2\ell +1) f^{(1)}_\ell(p) 
P_\ell(\cos \vartheta)\, ,
\end{equation}
where $\vartheta$ is the polar angle of ${\bf n}$, $\cos \vartheta ={\bf n} 
\cdot {\bf \hat{v}}$.

In the following we compute the second-order collision term separately for 
the different 
contributions, using the notation $C(\bp)=C^{(1)}(\bp)+C^{(2)}(\bp)/2$. 
We have not reported the details of the calculation of the first-order 
term because 
for its second-order analog, $c^{(2)}_{\Delta}(\bp,\bpp)+c^{(2)}_v(\bp,\bpp)$, 
the procedure is the same. The important 
difference is that the second-order velocity term includes a vector part, 
and this leads to 
a generic angular decomposition of the distribution function (for simplicity 
drop the time 
dependence) 
\begin{equation}
\label{fangdeco}
f^{(i)}({\bf x},p,{\bf n})=\sum_{\ell} \sum_{m=-\ell}^{\ell} 
f^{(i)}_{\ell m}({\bf x},p)  
(-i)^{\ell}\ \sqrt{\frac{4\pi}{2\ell+1}} Y_{\ell m}({\bf n})\, ,
\end{equation} 
such that 
\begin{equation}
\label{angular}
f^{(i)}_{\ell m}=(-i)^{- \ell}\sqrt{\frac{2\ell+1}{4\pi}} \int d\Omega  
f^{(i)} 
Y^{*}_{\ell m}({\bf n}) \, .
\end{equation}
Such a decomposition holds also in Fourier space \cite{Complete}.
The notation at this stage is a bit confusing, so let us restate it:
superscripts 
denote the order of the perturbation; the subscripts refer to the moments
of the distribution.  Indeed at first order one can drop the dependence on $m$ 
setting $m=0$ using the fact that the distribution function does not depend 
on the azimuthal angle 
$\phi$. In this case the relation with $f^{(1)}_l$ is 
\begin{equation}
\label{rel}
f^{(1)}_{\ell m}=(-i)^{-\ell} (2\ell +1)  \delta_{m0} \, f^{(1)}_{\ell}\, .
\end{equation} 

\noindent
a) $c^{(2)}_\Delta(\bp, \bpp)$:

The integral over $\bpp$ yields
\begin{eqnarray}
C^{(2)}_\Delta(\bp) &=&\frac{3n_e \sigma_T}{4p}\int dp' p' 
\frac{d \Omega'}{4 \pi} c^{(2)}_\Delta(\bp, \bpp) =
\frac{3n_e \sigma_T}{4p}\int dp' p'\delta(p-p') \nonumber \\
&\times & \int \frac{d \Omega'}{4 \pi} 
[1+({\bf n} \cdot {\bf n}')^2] [f^{(2)}(\bpp)-f^{(2)}(\bp)]\, . 
%\nonumber \\
\end{eqnarray}
To perform the angular integral we write 
the angular dependence on the scattering angle 
$\cos \theta= {\bf n} \cdot {\bf n}'$ in terms of the Legendre polynomials  
\begin{eqnarray}
\label{DL}
[1+({\bf n} \cdot {\bf n}')^2]&=&\frac{4}{3}\left[1+\frac{1}{2} 
P_2({\bf n} \cdot {\bf n}') \right] \nonumber \\
&=& 
\left[1+\frac{1}{2}\sum_{m=-2}^{2} Y_{2m}({\bf n})  Y^{*}_{2m}({\bf n}') 
\frac{4 \pi}{2\ell +1}
\right] \, ,
\end{eqnarray}
where in the last step we used the addition theorem for spherical harmonics
\begin{equation}
P_\ell=\frac{4 \pi}{2\ell +1} \sum_{m=-2}^{2} Y_{\ell m}({\bf n})  
Y^{*}_{\ell m}({\bf n}')\, . 
\end{equation}
Using the decomposition~(\ref{angular}) and the orthonormality of the 
spherical harmonics we find
\begin{eqnarray}
C^{(2)}_\Delta(\bp)=n_e \sigma_T \left[ 
f^{(2)}_{0 0}(p)-f^{(2)}(\bp)-\frac{1}{2} \sum_{m=-2}^{2} 
\frac{\sqrt{4 \pi}}{5^{3/2}}\, f^{(2)}_{2m}(p) \, Y_{2m}({\bf n}) 
\right]. \nonumber \\ 
\end{eqnarray} 
It is easy to recover the result for the corresponding first-order 
contribution in Eq.~(\ref{C1})
by using Eq.~(\ref{rel}).
\\
\noindent
b) $c^{(2)}_v(\bp,\bpp)$:

Let us fix for simplicity our coordinates such that the polar angle 
of ${\bf n}'$ is defined by $\mu'=
{\bf \hat{v}}^{(2)} \cdot {\bf n}'$ with $\phi'$ the corresponding 
azimuthal angle. The contribution 
of  $c^{(2)}_v(\bp,\bpp)$ to the collision term is then 
\begin{eqnarray}
C^{(2)}_v(\bp) &= &\frac{3 n_e \sigma_T}{4 p} v^{(2)} \int dp' p' 
[f^{(0)}(p')-f^{(0)}(p)]\frac{\partial \delta(p-p')}{\partial p'} 
\nonumber \\
&\times &\int_{-1}^{1} \frac{d \mu'}{2} (p \mu-p'\mu') \int_0^{2\pi} 
\frac{d \phi'}{2\pi} 
[1+({\bf p}\cdot {\bf p'})^2]\, .
\end{eqnarray} 
We can use Eq.~(\ref{DL}) which in our coordinate system reads
\begin{equation}
\label{DL2}
\frac{4}{3}\left[1+\frac{1}{2}  
\sum_{m=-2}^m \frac{(2-m)!}{(2+m)!} P_2^m({\bf n}\cdot{\bf 
\hat{ v}}^{(2)}) P_2^m({\bf n}' \cdot{\bf\hat{ v}}^{(2)})
e^{im(\phi'-\phi)} \right] \, ,
\end{equation}
so that 
\begin{equation}
\label{intphi}
\int \frac{d \phi'}{2 \pi} P_2({\bf n} \cdot{\bf n}') = 
P_2({\bf n} \cdot{\bf {\hat v}}^{(2)}) 
 P_2({\bf n}' \cdot{\bf {\hat v}}^{(2)}) =P_2(\mu)P_2(\mu')\, .
\end{equation}
By using the orthonormality of the Legendre polynomials and 
integrating by parts over $p'$ we find 
\begin{equation}
C^{(2)}_v(\bp)= - n_e\,  \sigma_T\,
p \frac{\partial f^{(0)}}{\partial p} {\bf v}^{(2)} \cdot {\bf n}\, . 
\end{equation}
As it is clear by the presence of the scalar product ${\bf v}^{(2)} 
\cdot {\bp}$ the final result is 
independent of the coordinates chosen.
\\
\noindent
c) $c^{(2)}_{\Delta v}(\bp,\bpp)$:

Let us consider the contribution from the first term
$$
c^{(2)}_{\Delta v(I)}(\bp,\bpp)=\left(1+\cos^2\theta\right) 
\left(\fps1 - \f1\right)
(\bp-\bpp)\cdot\bv {\partial \delta(p-p')
			\over \partial p'}\, ,  
$$
where the velocity has to be considered at first order. 
In the integral~(\ref{Integralcolli}) it brings
\begin{eqnarray}
\frac{1}{2}C^{(2)}_{\Delta v (I)} & = & \frac{3 n_e \sigma_T v}{4 p} 
\int dp' p' 
\frac{\partial \delta(p-p')}{\partial p'} 
\int_{-1}^1 \frac{d \mu'}{2} [f^{(1)}(\bpp)-f^{(1)}(\bp)] \nonumber \\
&\times &(p\mu - p'\mu') 
\int_0^{2\pi} \frac{d \phi'}{2 \pi} (1+\cos^2 \theta) \, ,
\end{eqnarray} 
The procedure to do the integral is the same as above. We use the 
same relations as in
Eqs.~(\ref{DL2}) and~(\ref{intphi}) where now the angles are those taken 
with respect to 
the first-order velocity. This eliminates the integral over $\phi'$, and 
integrating by parts over $p'$ yields
\begin{eqnarray}
\label{interm}
&&\frac{1}{2} 
C^{(2)}_{\Delta v (I)}(\bp)=-\frac{3 n_e \sigma_T v}{4p} \int_{-1}^1 
\frac{d \mu'}{2} 
\left[
\frac{4}{3}+\frac{2}{3} P_2(\mu) P_2(\mu') 
\right] \\
&&\times 
\biggl[ 
p(\mu-2\mu')(f^{(1)}(p,\mu')-f^{(1)}(p,\mu))
+p^2(\mu-\mu') \frac{\partial f^{(1)}(p,\mu')}{\partial p} 
\biggr]\, . \nonumber
\end{eqnarray} 
We now  use the decomposition~(\ref{dec1}) and the orthonormality of the 
Legendre polynomials to find
\begin{eqnarray}
&&\int \frac{d\mu'}{2} \mu' f^{(1)}(p,\mu') P_2(\mu')=
\sum_\ell \int \frac{d\mu'}{2} \mu ' P_2(\mu') P_l(\mu') f^{(1)}_\ell(p)
\nonumber \\
&&=
\sum_\ell \int \frac{d\mu'}{2} \left[ \frac{2}{5} P_1(\mu') +\frac{3}{5} 
P_3(\mu') \right] 
P_\ell(\mu')f^{(1)}_\ell(p) \nonumber \\
&&= \frac{2}{5} f^{(1)}_1(p)+\frac{3}{5} f^{(1)}_3(p)\, ,
\end{eqnarray}
where we have used $\mu ' P_2(\mu') P_l(\mu')=\frac{2}{5} P_1(\mu') +
\frac{3}{5} P_3(\mu')$, with 
$P_1(\mu')=\mu'$. Thus from Eq.~(\ref{interm}) we get 
\begin{eqnarray}
\frac{1}{2} C^{(2)}_{\Delta v (I)}(\bp)&=& n_e \sigma_T \Bigg\{ 
{\bf v} \cdot {\bf n} \biggl[ 
f^{(1)}(\bp) -f^{(1)}_0(p) - p\frac{\partial f^{(1)}_0(p)}{\partial p} 
\nonumber \\
&-& \frac{1}{2} P_2({\bf \hat{v}} \cdot {\bf n}) \left( f^{(1)}_2(p)+
p\frac{\partial f^{(1)}_2(p)}{\partial p}
\right) \biggr] \nonumber \\
&+& v \biggl[ 2f^{(1)}_1(p)+ p\frac{\partial f^{(1)}_1(p)}{\partial p} + 
\frac{1}{5} 
P_2({\bf \hat{v}}\cdot {\bf n}) \biggl( 2f^{(1)}_1(p) \nonumber \\
&+& p\frac{\partial 
f^{(1)}_1(p)}{\partial p}+
3 f^{(1)}_3(p)+\frac{3}{2} p\frac{\partial f^{(1)}_3(p)}{\partial p}
\biggr)
\biggr] \Bigg\} \, .
\end{eqnarray}
In $c^{(2)}(\bp,\bpp)$ there is a second term
$$c^{(2)}_{\Delta v(II)}= 
-2\cos\theta(1-\cos\theta) 
\left( f^{(1)}(\bpp)-f^{(1)}(\bp) \right)
\delta(p-p') ({\bf n}+ {\bf n}')\cdot \bv, $$
whose contribution to the collision term is 
\begin{eqnarray}
\frac{1}{2}C^{(2)}_{\Delta v (II)}(\bp) &=& -\frac{3n_e \sigma_T v}{2p} 
\int dp' p' \delta(p-p') \int_{-1}^1 
\frac{d\mu'}{2} (f^{(1)}(\bpp) \nonumber \\
&-& f^{(1)}(\bp)) (\mu+\mu')\int_0^{2\pi} 
\frac{d \phi'}{2\pi} 
\cos \theta(1-\cos\theta)\, .
\end{eqnarray}
This integration proceeds through the same steps as for 
$C^{(2)}_{\Delta v (I)}(\bp)$. In particular
by noting that $\cos \theta(1-\cos \theta)=-1/3+P_1(\cos \theta)
-2P_3(\cos\theta)/3$, 
Eqs.~(\ref{DL2}) and~(\ref{intphi}) allows to compute    
\begin{equation}
\int \frac{d \phi'}{2 \pi} \cos \theta (1-\cos \theta)=
-\frac{1}{3} +P_1(\mu) P_1(\mu')-\frac{2}{3} P_2(\mu) P_2(\mu')\, , 
\end{equation} 
and using the decomposition~(\ref{dec1}) we arrive at
\begin{eqnarray}
\label{interm2}
\frac{1}{2}C^{(2)}_{\Delta v(II)}(\bp) &=& - n_e \sigma_T 
\biggl\{ {\bf v} \cdot {\bf n}\, f^{(1)}_2(p) (1-P_2({\bf {\hat v}}
\cdot {\bf n})) \nonumber \\
&+& v \left[\frac{1}{5} \,
P_2({\bf {\hat v}}\cdot {\bf n})\,  \left( 
 3 f^{(1)}_1(p)-3 f^{(1)}_3(p) \right)
\right]
\biggr\}\, .
\end{eqnarray}
We then obtain
\begin{eqnarray}
\frac{1}{2}C^{(2)}_{\Delta v}(\bp) &=&
n_e \sigma_T \Bigg\{ {\bf v} \cdot {\bf n} \biggl[ 
f^{(1)}(\bp) -f^{(1)}_0(p) - p\frac{\partial f^{(1)}_0(p)}{\partial p} 
-f^{(1)}_2(p) \nonumber \\
&+ &\frac{1}{2} P_2({\bf \hat{v}} \cdot {\bf n}) \biggl( f^{(1)}_2(p)-
p\frac{\partial f^{(1)}_2(p)}{\partial p}
\biggr) \biggr] + v \biggl[ 2f^{(1)}_1(p) \nonumber \\ 
&+& p\frac{\partial f^{(1)}_1(p)}{\partial p} + 
\frac{1}{5} P_2({\bf \hat{v}}\cdot {\bf n}) 
\biggl( -f^{(1)}_1(p) \nonumber \\
&+& p\frac{\partial 
f^{(1)}_1(p)}{\partial p}+
6 f^{(1)}_3(p)+\frac{3}{2} p\frac{\partial f^{(1)}_3(p)}{\partial p}
\biggr)
\biggr]
\Bigg\} \, .
\end{eqnarray}
As far as the remaining terms, these have already 
been computed in Ref.~\cite{DJ} (see also 
Ref.~\cite{HSS})  and here we just report them
\\

\noindent
d) $c^{(2)}_{v v}(\bp,\bpp)$:

The term proportional to the velocity squared 
yield a contribution to the collision term
\begin{eqnarray}
\frac{1}{2}C^{(2)}_{v v}(\bp) &=& 
n_e \sigma_T \biggl\{ ({\bf v} \cdot 
{\bf n})^2 \left[ 
p \frac{\partial f^{(0)}}{\partial p}+\frac{11}{20} p^2 
\frac{\partial^2 f^{(0)}}{\partial p^2} \right] \nonumber \\
&+& v^2 \left[ 
p \frac{\partial f^{(0)}}{\partial p}+\frac{3}{20} p^2 
\frac{\partial^2 f^{(0)}}{\partial p^2}
\right]
\biggr\}\, .
\end{eqnarray}

\noindent
e) $c^{(2)}_K(\bp,\bp')$: 

The terms responsible for the spectral distortions give
\begin{equation}
\frac{1}{2}C_K^{(2)}(\bp)=\frac{1}{m_e^2}\frac{\partial}{\partial p}\left\{ 
p^4\left[
T_e \frac{\partial f^{(0)}}{\partial p}+f^{(0)}(1+f^{(0)})
\right]
\right\}\, .
\end{equation}
Finally, we write also the part of the collision term coming from 
Eq.~(\ref{deltac1})
\begin{eqnarray}
\delta^{(1)}_e\, c^{(1)}(\bp,\bpp) &\rightarrow &\delta^{(1)}_e C^{(1)}(\bp) =
 n_e \sigma_T\, \delta^{(1)}_e \biggl[ f^{(1)}_0(p) \nonumber \\
&+& \frac{1}{2}f^{(1)}_2 
P_2({\bf \hat v} \cdot 
{\bf n})-f^{(1)}-p\frac{\partial f^{(0)}}{\partial p} {\bf v} \cdot {\bf n} 
\biggr]\, . 
\end{eqnarray}

%%%%%%%%%%%%%%%%%%%%%%%%%%%%%%%%%%%%%%%%%%%%%%%%%%%%%%%%%%%%%%%%%%%%%%%%%
\subsection{Final expression for the collision term}
%%%%%%%%%%%%%%%%%%%%%%%%%%%%%%%%%%%%%%%%%%%%%%%%%%%%%%%%%%%%%%%%%%%%%%%%%

Summing all the terms we find the final expression for the collision 
term~(\ref{Integralcolli}) up to second order
\begin{eqnarray}
C(\bp)=C^{(1)}(\bp)+\frac{1}{2} C^{(2)}(\bp)
\end{eqnarray}
with 
\begin{eqnarray}
\label{C1p}
C^{(1)}(\bp)= n_e \sigma_T \left[ f^{(1)}_0(p)+\frac{1}{2}f^{(1)}_2 P_2({\bf 
\hat v} \cdot 
{\bf n})-f^{(1)}-p\frac{\partial f^{(0)}}{\partial p} {\bf v} \cdot {\bf n} 
\right]
\end{eqnarray}
and 
\begin{eqnarray}
\label{C2p}
\frac{1}{2}C^{(2)}(\bp)&=&n_e \sigma_T \Bigg\{ 
\frac{1}{2}f^{(2)}_{0 0}(p)-\frac{1}{4} \sum_{m=-2}^{2} 
\frac{\sqrt{4 \pi}}{5^{3/2}}\, f^{(2)}_{2m}(p) \, Y_{2m}({\bf n}) \nonumber \\
&-&\frac{1}{2} f^{(2)}(\bp) 
+ \delta^{(1)}_e \biggl[ f^{(1)}_0(p)+\frac{1}{2}f^{(1)}_2 P_2({\bf \hat v} 
\cdot {\bf n})-f^{(1)} \nonumber \\
&-& p\frac{\partial f^{(0)}}{\partial p} {\bf v} \cdot {\bf n} 
\biggr] 
- \frac{1}{2}p\frac{\partial f^{(0)}}{\partial p} {\bf v}^{(2)} 
\cdot {\bf n}+ {\bf v} \cdot {\bf n} \biggl[ f^{(1)}(\bp) \nonumber \\
&-& f^{(1)}_0(p) - p\frac{\partial f^{(1)}_0(p)}{\partial p} 
-f^{(1)}_2(p) 
+ \frac{1}{2} P_2({\bf \hat{v}} \cdot {\bf n}) \nonumber \\
&\times & \left( f^{(1)}_2(p)-
p\frac{\partial f^{(1)}_2(p)}{\partial p}
\right) \biggr] 
+ v \biggl[ 2f^{(1)}_1(p)+ p\frac{\partial f^{(1)}_1(p)}{\partial p} 
\nonumber \\
&+ & \frac{1}{5} 
P_2({\bf \hat{v}}\cdot {\bf n}) \biggl( - f^{(1)}_1(p)+ p\frac{\partial 
f^{(1)}_1(p)}{\partial p}+
6 f^{(1)}_3(p) \nonumber \\
&+& \frac{3}{2} p\frac{\partial f^{(1)}_3(p)}{\partial p}
\biggr) \biggr] + ({\bf v} \cdot {\bf n})^2 \biggl[ 
p \frac{\partial f^{(0)}}{\partial p}+\frac{11}{20} p^2 
\frac{\partial^2 f^{(0)}}{\partial p^2}
\biggr] \nonumber \\
&+&v^2 \biggl[ 
p \frac{\partial f^{(0)}}{\partial p}+\frac{3}{20} p^2 
\frac{\partial^2 f^{(0)}}{\partial p^2}
\biggr] \nonumber \\
&+& \frac{1}{m_e^2}\frac{\partial}{\partial p}\biggl[
p^4\biggl(
T_e \frac{\partial f^{(0)}}{\partial p}+f^{(0)}(1+f^{(0)})
\biggr) \biggr] \Bigg\}\, .
\end{eqnarray}
Notice that there is an internal hierarchy, with terms which do not depend 
on the baryon velocity 
${\bf v}$, terms proportional to ${\bf v} \cdot {\bf n}$ and then to 
$({\bf v} \cdot {\bf n})^2$, 
$v$ and $v^2$ (apart from the Kompaneets terms). In particular notice the 
term  proportional to 
$\delta^{(1)}_e {\bf v} \cdot {\bf n}$ is the one corresponding to the 
Vishniac effect. 
We point out that we have kept all the 
terms up to second order in the collision term. In Refs.~\cite{DJ,HSS} many 
terms coming from $c^{(2)}_{\Delta v}$ 
have been dropped mainly because these terms are proportional to the photon 
distribution function $f^{(1)}$ which on 
very small scales (those of interest for reionization) is suppressed by the 
diffusion damping. Here we want 
to be completely general and we have to keep them.   

%%%%%%%%%%%%%%%%%%%%%%%%%%%%%%%%%%%%%%%%%%%%%%%%%%%%%%%%%%%%%%%%%%%%%%%%%%%
\section{The Brightness equation}
%%%%%%%%%%%%%%%%%%%%%%%%%%%%%%%%%%%%%%%%%%%%%%%%%%%%%%%%%%%%%%%%%%%%%%%%%%%%
 
\subsection{First order}

The Boltzmann equation for photons is obtained by combining 
Eq.~(\ref{LHSBoltz}) with Eqs.~(\ref{C1p})-(\ref{C2p}). 
At first order the left-hand side reads
\begin{eqnarray}
\label{Bdf1}
\frac{df}{d\eta}&=&\frac{d f^{(1)}}{d \eta}-p
\frac{\partial f^{(0)}}{\partial p} \frac{\partial \Phi^{(1)}}{\partial x^i} 
\frac{dx^i}{d\eta}+p\frac{\partial f^{(0)}}{\partial p} 
\frac{\partial \Psi^{(1)}}{ \partial \eta}\, .
\end{eqnarray}
At first order it is useful to characterize the perturbations to 
the Bose-Einstein distribution function~(\ref{BEd}) 
in terms of a perturbation to the temperature as 
\begin{equation}
\label{ft}
f(x^i,p,n^i,\eta)=2\, \left[ \exp\left\{ \frac{p}{T(\eta)
(1+ \Theta^{(1)})}\right\}-1 \right]^{-1}\, .
\end{equation}  
Thus it turns out that 
\begin{equation}
\label{thetaf1}
f^{(1)}=-p\frac{\partial f^{(0)}}{\partial p} \Theta^{(1)}\, ,
\end{equation} 
where we have used the fact that $\partial f/\partial \Theta|_{\Theta=0}=-p 
\partial f^{(0)}/\partial p$. In terms of this variable $\Theta^{(1)}$   
the linear collision term~(\ref{C1p}) will now become proportional to 
$- p \partial f^{(0)}/\partial p$ which contains the only explicit 
dependence on $p$, and the same happens for the left-hand side, 
Eq.~(\ref{Bdf1}). This is telling us that at first order $\Theta^{(1)}$ 
does not depend on $p$ but only on ${x^i, n^i, \eta}$,  
$\Theta^{(1)}=\Theta^{(1)}({x^i, n^i, \tau})$. This is well known 
and the physical reason is that at linear order there is no energy 
transfer in Compton collisions between photons and electrons. Therefore, 
the Boltzmann equation for $\Theta^{(1)}$ reads
\begin{eqnarray}
\label{BoltzTheta}
&&\frac{\partial \Theta^{(1)}}{\partial \eta}+n^i \frac{\partial 
\Theta^{(1)}}{\partial x^i}
+\frac{\partial \Phi^{(1)}}{\partial x^i} n^i -\frac{\partial 
\Psi^{(1)}}{ \partial \eta} \nonumber \\
&& = 
n_e\sigma_T a \left[ \Theta^{(1)}_0+\frac{1}{2} \Theta^{(1)}_2 
P_2({\bf {\hat v}} \cdot {\bf n})-\Theta^{(1)}+{\bf v} \cdot {\bf n} 
\right]\, ,
\end{eqnarray} 
where we made us of 
$f^{(1)}_{\ell}=- p \partial f^{(0)}/\partial p 
\Theta^{(1)}_\ell$, according to the decomposition of 
Eq.~(\ref{dec1}), and we have 
taken the zero-order expressions for $dx^i/d\eta$, dropping the 
contribution from $dn^i/d\eta$ in Eq.~(\ref{Df}) since it is already 
first-order. 

Notice that, since $\Theta^{(1)}$ is independent of $p$, it is 
equivalent to consider the quantity
\begin{equation}
\label{Delta1}
\Delta^{(1)}(x^i,n^i,\tau)=
\frac{\int dp p^3 f^{(1)}}{\int dp p^3 f^{(0)}}\, , 
\end{equation}
being $\Delta^{(1)}=4\Theta^{(1)}$ at this order.  
The physical meaning of $\Delta^{(1)}$ is that of a fractional 
energy perturbation (in a given direction). From Eq.~(\ref{LHSBoltz}) another 
way to write an equation for $\Delta^{(1)}$ -- the so-called 
brightness equation -- is     
\begin{eqnarray}
\label{B1}
&&\frac{d}{d\eta} \left[ \Delta^{(1)}+4 \Phi^{(1)} \right]-4 
\frac{\partial}{\partial \eta}\left( \Phi^{(1)}+\Psi^{(1)} \right) 
\nonumber \\
&& = n_e \sigma_T a \left[ \Delta^{(1)}_0+\frac{1}{2} \Delta^{(1)}_2 
P_2({\bf {\hat v}} \cdot {\bf n})-\Delta^{(1)}+4 {\bf v} \cdot {\bf n}
\right]\, .
\end{eqnarray}
%%%%%%%%%%%%%%%%%%%%%%%%%%%%%%%%%%%%%%%%%%%%%%%%%%%%%%%%%%%%%%%%%%%%%%%%
\subsection{Second order}
%%%%%%%%%%%%%%%%%%%%%%%%%%%%%%%%%%%%%%%%%%%%%%%%%%%%%%%%%%%%%%%%%%%%%
 
The previous results show that at linear order the photon distribution 
function has a Planck spectrum with the temperature that at any point 
depends on the photon direction. At second order one could characterize the 
perturbed 
photon distribution function in a similar way as in Eq.~(\ref{ft})  
\begin{equation}
f(x^i,p,n^i,\eta)=2\, \left[ \exp\left\{ \frac{p}{T(\eta)\, e^{\Theta}}-1 
\right\}\right]^{-1}\, ,
\end{equation}
where by expanding $\Theta=\Theta^{(1)}+\Theta^{(2)}/2+...$ as usual one 
recovers the first-order expression. 
For example, in terms of $\Theta$, the perturbation of $f^{(1)}$ is given 
by Eq.~(\ref{thetaf1}), while at second order  
\begin{eqnarray}
\frac{f^{(2)}}{2}
=-\frac{p}{2} \frac{\partial f^{(0)}}{\partial p} \Theta^{(2)}+\frac{1}{2} 
\left(p^2 \frac{\partial^2f^{(0)}}{\partial p^2}
+ p \frac{\partial f^{(0)}}{\partial p}  \right) \left( \Theta^{(1)} 
\right)^2\, .
\end{eqnarray}
However, as discussed in details in Refs.~\cite{HSS,DJ}, now the second-order 
perturbation $\Theta^{(2)}$ will not be momentum independent 
because the collision term in the equation for $\Theta^{(2)}$ does depend 
explicitly on $p$ (defining the combination  
$- (p \partial f^{(0)}/\partial p)^{-1} f^{(2)}$ does not lead to a 
second-order momentum independent equation as above). Such dependence 
is evident, for example, in the terms of $C^{(2)}({\bf p})$, Eq.~(\ref{C2p}), 
proportional to $v$ or $v^2$, and in the Kompaneets terms. 
The physical reason is that at the non-linear level photons and electrons do 
exchange energy during Compton collisions. As a consequence 
spectral 
distorsions are generated. For example, in the isotropic limit,
 only the Kompaneets terms survive giving rise to the Sunyaev-Zeldovich 
distorsions. As discussed in Ref.~\cite{HSS}, the Sunyaev-Zeldovich 
distorsions can also be obtained with the correct coefficients by replacing 
the average over the direction electron 
$\langle v^2 \rangle$ with the mean squared thermal velocity $\langle 
v_{th}^2 \rangle=3T_e/m_e$ in Eq.~(\ref{C2p}). 
This is due simply to the fact that 
the distinction between thermal and bulk velocity of the electrons is 
just for convenience. This fact also 
shows that spectral distorsions due 
to the bulk flow (kinetic Sunyaev-Zeldovich) has the same form as the 
thermal effect. Thus 
spectral distorsions can be in general described by a global Compton 
$y$-parameter 
(see Ref.~\cite{HSS} for a full discussion of spectral distorsions). 
However in the following we will not be 
interested in the frequency dependence but only in the anisotropies of 
the radiation distribution. Therefore we can integrate over the 
momentum 
$p$ and define ~\cite{HSS,DJ} 
\begin{equation}
\label{Delta2}
\Delta^{(2)}(x^i,n^i,\tau)=\frac{\int dp p^3 f^{(2)}}{\int dp p^3 f^{(0)}}\, ,
\end{equation}
as in Eq.~(\ref{Delta1}).

Integration over $p$ of Eqs.~(\ref{LHSBoltz})-(\ref{C2p}) is straightforward 
using the following relations
\begin{eqnarray}
\label{rules}
\int dp p^3 p \frac{\partial f^{(0)}}{\partial p}&=&-4\, N\, ; \quad \quad
\int dp p^3 p^2 \frac{\partial^2 f^{(0)}}{\partial p^2}= 20\, N\, ; 
\nonumber\\
\int dp p^3 f^{(1)} &=&N \Delta^{(1)} \, ; \quad \quad
\int dp p^3 p \frac{\partial f^{(1)}}{\partial p}=-4\, N \Delta^{(1)}\, .
\end{eqnarray}
Here $N=\int dp p^3 f^{(0)}$ is the normalization factor
(it is just proportional the background energy density of 
photons ${\bar \rho}_\gamma$). At first order one recovers Eq.~(\ref{B1}). 
At second order we find
\begin{eqnarray}
\label{B2}
& & \frac{1}{2} \frac{d}{d\eta} \left[ \Delta^{(2)}+4\Phi^{(2)} \right] + 
\frac{d}{d\eta} \left[ \Delta^{(1)} +4 \Phi^{(1)} \right] 
-4 \Delta^{(1)}\left( \Psi^{(1)'}-\Phi^{(1)}_{,i}n^i \right) \nonumber \\
&&- 2 
\frac{\partial}{\partial \eta}\left( \Psi^{(2)}+\Phi^{(2)} \right) 
+4 \frac{\partial \omega_i}{\partial \eta} n^i + 2 \frac{\partial 
\chi_{ij}}{\partial \eta} n^i n^j = \nonumber \\
& &= - \frac{\tau'}{2} \Bigg[ \Delta^{(2)}_{00} -\Delta^{(2)} 
- \frac{1}{2} \sum_{m=-2}^{2} \frac{\sqrt{4 \pi}}{5^{3/2}}\, 
\Delta^{(2)}_{2m} \, Y_{2m}({\bf n}) +
2 (\delta^{(1)}_e +\Phi^{(1)}) \nonumber \\
&& \left( \Delta^{(1)}_0+\frac{1}{2} \Delta^{(1)}_2 P_2({\bf {\hat v}} 
\cdot {\bf n})-\Delta^{(1)}+4 {\bf v} \cdot {\bf n}
\right)+4{\bf v}^{(2)} \cdot {\bf n} \nonumber \\
& & + 2 ({\bf v} \cdot {\bf n}) \left[ \Delta^{(1)}+3\Delta^{(1)}_0-
\Delta^{(1)}_2 \left(1-\frac{5}{2} P_2({\bf {\hat v}} 
\cdot {\bf n}) \right)\right] \nonumber \\
&& - v\Delta^{(1)}_1 \left(4+2 P_2({\bf {\hat v}} \cdot {\bf n}) 
\right) 
+14 ({\bf v} \cdot {\bf n})^2-2 v^2  \Bigg]\, ,
\end{eqnarray}
where we have expanded the angular dependence of $\Delta$ as in 
Eq.~(\ref{fangdeco})
\begin{equation}
\label{Dlm2}
\Delta^{(i)}({\bf x}, {\bf n})=\sum_{\ell} \sum_{m=-\ell}^{\ell} 
\Delta^{(i)}_{\ell m}({\bf x})  
(-i)^{\ell}  \sqrt{\frac{4 \pi}{2\ell+1}}Y_{\ell m}({\bf n})\, ,
\end{equation} 
with  
\begin{equation}
\label{angular1}
\Delta^{(i)}_{\ell m}=(-i)^{- \ell}\sqrt{\frac{2\ell+1}{4\pi}} 
\int d\Omega  \Delta^{(i)} 
Y^{*}_{\ell m}({\bf n}) \, ,
\end{equation}
where we recall that the superscript stands by the order of the 
perturbation. At first order one can drop the dependence on $m$ 
setting $m=0$ so that  $\Delta^{(1)}_{\ell m}=(-i)^{-\ell} (2\ell +1)  
\delta_{m0} \, \Delta^{(1)}_{\ell}$. In Eq.~(\ref{B2}) we have 
introduced the differential optical depth 
\begin{equation}
\label{deftau}
\tau'=-{\bar n}_e \sigma_T a \, .
\end{equation}
It is understood that
 on the left-hand side of Eq.~(\ref{B2}) one has to 
pick up for the total time derivatives only those terms which contribute to 
second order. Thus we have to take 
\begin{eqnarray}
&&\frac{1}{2} \frac{d}{d\eta} \left[ \Delta^{(2)}+4\Phi^{(2)} \right] + 
\frac{d}{d\eta} \left[ \Delta^{(1)} +4 \Phi^{(1)} \right]\Big|^{(2)} \\
&&= \frac{1}{2}\left( \frac{\partial}{\partial \eta}+n^i
 \frac{\partial}{\partial x^i}\right) \left( \Delta^{(2)}+4\Phi^{(2)} \right)
+n^i(\Phi^{(1)}+\Psi^{(1)}) \nonumber \\
&& \times 
\partial_i(\Delta^{(1)}+4\Phi^{(1)}) + 
\left[(\Phi^{(1)}_{,j}+\Psi^{(1)}_{,j})n^in^j 
-(\Phi^{,i}+\Psi^{,i})\right]
\frac{\partial \Delta^{(1)}}{\partial n^i}\, , \nonumber
\end{eqnarray}
where we used Eqs.~(\ref{dxi}) and ~(\ref{dni}). Notice that we can
 write $\partial \Delta^{(1)}/\partial n^i=
(\partial \Delta^{(1)}/\partial x^i) (\partial x^i /\partial n^i)=
(\partial \Delta^{(1)}/\partial x^i) (\eta-\eta_i)$, from the 
integration in time of Eq.~(\ref{dxi}) at zero-order when $n^i$ is 
constant in time. 

%%%%%%%%%%%%%%%%%%%%%%%%%%%%%%%%%%%%%%%%%%%%%%%%%%%%%%%%%%%%%%%%%%%%%%%%%%%
\subsection{Hierarchy equations for multipole moments}
%%%%%%%%%%%%%%%%%%%%%%%%%%%%%%%%%%%%%%%%%%%%%%%%%%%%%%%%%%%%%%%%%%%%%%%%%

Let us now move to Fourier space. In the following, for a given wave-vector 
${\bf k}$ we will choose the coordinate system such that 
${\bf e}_3={\bf {\hat k}}$ and the polar angle 
of the photon momentum is $\vartheta$, with $\mu=\cos \vartheta = 
{\bf \hat{k}} \cdot {\bf n}$. Then Eq.~(\ref{B2}) can be written as 
\begin{equation}
\label{BF}
\Delta^{(2) \prime}+ik\mu \Delta^{(2)}-\tau' \Delta^{(2)}= 
S({\bf k},{\bf n},\eta)\, ,
\end{equation}
where $S({\bf k},{\bf n},\eta)$ can be easily read off Eq.~(\ref{B2}). 
We now expand the temperature anisotropy in the multipole moments 
$\Delta^{(2)}_{\ell m}$ in order to obtain a system of coupled 
differential equations. By applying the angular integral of 
Eq.~(\ref{angular1}) to Eq.~(\ref{BF}) we find
\begin{equation}
\label{H}
\Delta^{(2)\prime }_{\ell m}({\bf k},\eta)=k 
\left[\frac{{\kappa}_{\ell m}}{2\ell-1}\Delta^{(2)}_{\ell-1,m}-
\frac{{\kappa}_{\ell+1, m}}{2\ell+3}\Delta^{(2)}_{\ell+1,m} \right] + 
\tau' \Delta^{(2)}_{\ell m}+ S_{\ell m}
\end{equation}  
where the expansion coefficients of the source term are given by
\begin{eqnarray}
\label{Slm}
S_{\ell m}&=& \left( 4 \Psi^{(2)\prime }-\tau' \Delta^{(2)}_{00} \right) 
\delta_{\ell0} \delta_{m0} +4 k \Phi^{(2)} 
\delta_{\ell 1} \delta_{m 0}-4\, \omega^\prime_{\pm 1} 
\delta_{\ell 1} \nonumber \\
&-& 8 \tau' 
v^{(2)}_m \delta_{\ell 1}-\frac{\tau'}{10} 
\Delta^{(2)}_{\ell m} \,\delta_{\ell 2} -2\chi^\prime_{\pm 2} 
\,\delta_{\ell 2} 
\nonumber \\
&-&2 \tau' \int \frac{d^3k_1}{(2\pi)^3} \biggl[ v^{(1)}_0({\bf k}_1) 
v^{(1)}_0({\bf k}_2) {\bf {\hat k}}_2 \cdot  
{\bf {\hat k}}_1  + (\delta^{(1)}_e({\bf k}_1)+\Phi^{(1)}({\bf k}_1)) 
\nonumber \\
&\times & \Delta^{(1)}_0({\bf k}_2) -i\frac{2}{3} 
v({\bf k}_1) \Delta^{(1)}_{10}({\bf k}_2) \biggr] \delta_{\ell 0} 
\delta_{m 0}\nonumber \\
&+& 16 k\, \int \frac{d^3k_1}{(2\pi)^3} \biggl[ \Phi^{(1)}({\bf k}_1) 
\Phi^{(1)}({\bf k}_2) 
\biggr] \delta_{\ell 1} \delta_{m 0} -2 \biggl[ (\Psi^{(1)} \nabla 
\Phi^{(1)})_m \nonumber \\
&+& 8\tau' [(\delta^{(1)}_e+\Phi) {\bf v}]_m +6 \tau' 
(\Delta^{(1)}_0 {\bf v})_m-2 \tau' (\Delta^{(1)}_2 {\bf v})_m  \biggr]\, 
\delta_{\ell 1} + \tau' \nonumber \\
&\times & \int \frac{d^3k_1}{(2\pi)^3} \left[ (\delta^{(1)}_e({\bf k}_1)+
\Phi^{(1)}({\bf k}_1)) \Delta^{(1)}_2({\bf k}_2)
-i\frac{2}{3}v({\bf k}_1) \Delta^{(1)}_{10}({\bf k}_2) \right]\, 
\nonumber \\
&\times & \delta_{\ell 2} \, \delta_{m 0}
+\int \frac{d^3k_1}{(2\pi)^3} \biggl[ 8 \Psi^{(1)}({\bf k}_1)
+ 2 \tau' (\delta^{(1)}_e({\bf k}_1)+\Phi^{(1)}({\bf k}_1)) \nonumber \\
&-& (\eta-\eta_i) (\Psi^{(1)}+\Phi^{(1)})({\bf k}_1) \, {\bf 
k}_1\cdot{\bf k}_2  \biggr] \Delta^{(1)}_{\ell0}({\bf k}_2)
\delta_{m0} - i (-i)^{-\ell} \nonumber \\
&\times & 
(-1)^{-m}(2\ell +1) \sum_{\ell''} 
\sum_{m'=-1}^{1}(2\ell''+1) \biggl[ 8 \Delta^{(1)}_{\ell''} \nabla \Phi^{(1)}
-2(\Phi^{(1)} \nonumber \\
&+&\Psi^{(1)}) \nabla \Delta^{(1)}_{\ell''}\biggr]_{m'}
\left(\begin{array}{ccc}\ell''&1&\ell\\0&0&0\end{array}\right)
\left(\begin{array}{ccc}\ell''&1&\ell\\0&m'&-m\end{array}\right) \nonumber \\
&+& \tau' i (-i)^{-\ell}(-1)^{-m} (2\ell +1) \sum_{\ell''} 
\sum_{m'=-1}^{1}(2\ell''+1) \biggl[ 2 \Delta^{(1)}_{\ell''} {\bf v} 
\nonumber \\
&+ &5  \delta_{\ell'' 2} \, \Delta^{(1)}_2 {\bf v} \biggr]_{m'}
\left(\begin{array}{ccc}\ell''&1&\ell\\0&0&0\end{array}\right)
\left(\begin{array}{ccc}\ell''&1&\ell\\0&m'&-m\end{array}\right) \nonumber \\
&+&14\tau'(-i)^{-\ell}(-1)^{-m} \sum_{m',m''=-1}^{1} 
\int \frac{d^3k_1}{(2\pi)^3} \biggl[ v^{(1)}_0({\bf k}_1) 
\frac{v^{(1)}_0({\bf k}_2)}{k_2} \nonumber \\
&\times & \frac{4\pi}{3} 
Y^*_{1m'}({\bf {\hat k}}_1) \left( k Y^*_{1m''}({\bf {\hat k}})-
 k_1 Y^*_{1m''}({\bf {\hat k}_1})\right)
\biggr]  \nonumber \\ 
&\times&\left(\begin{array}{ccc}1&1&\ell\\0&0&0\end{array}\right)
\left(\begin{array}{ccc}1&1&\ell\\m'&m''&-m\end{array}\right)
+2 (\eta-\eta_i)(-i)^{-\ell} \nonumber \\
&\times & (-1)^{-m} \sqrt{\frac{2\ell+1}{4\pi}} 
\sum_L \sum_{m',m''=-1}^{1} 
\int \frac{d^3k_1}{(2\pi)^3} 
\sqrt{\frac{4\pi}{2L+1}} 
\left( \frac{4\pi}{3} \right)^2 \nonumber \\
&\times&
\Delta^{(1)}_L({\bf k}_1) (\Phi^{(1)} + \Psi^{(1)})({\bf k}_2)
k_1 Y^*_{1m'}({\bf {\hat k}}_1)\left( k Y^*_{1m''}
({\bf {\hat k}}) \right. \nonumber \\
&-& \left.
k_1 Y^*_{1m''}({\bf {\hat k}_1})\right)
 \int d \Omega Y_{1m'}({\bf n}) Y_{1m''}({\bf n}) Y_{L0}({\bf n}) 
Y_{\ell-m}({\bf n})\, ,
\end{eqnarray}
where ${\bf k}_2={\bf k}-{\bf k}_1$ and $k_2=|{\bf k}_2|$. 
In Eq.~(\ref{Slm}) it is understood that $|m| \leq \ell$.

Let us explain the notations we have adopted in writing Eq.~(\ref{Slm}). 
The baryon velocity at linear order is irrotational, 
meaning that it is the gradient of a potential, and thus in Fourier space 
it is parallel to 
${\bf {\hat k}}$, and following the same 
notation of Ref.~\cite{husolo}, we write 
\begin{equation}
\label{vzero}
{\bf v}^{(1)}({\bf k})=-i v^{(1)}_0({\bf k}) {\bf {\hat k}}\, .
\end{equation}
The second-order velocity perturbation will contain a transverse 
(divergence-free) part whose components are orthogonal to  
${\bf {\hat k}}={\bf e}_3$, and we can write 
\begin{equation}
\label{vdec}
{\bf v}^{(2)}({\bf k})=-i v^{(2)}_0({\bf k}) {\bf e}_3+\sum_{m=\pm1}
 v^{(2)}_m \, \frac{{\bf e}_2\mp i {\bf e}_1}{\sqrt 2}\, ,
\end{equation}
where ${\bf e}_i$ form an orhtonormal basis with  ${\bf {\hat k}}$. 
The second-order
perturbation $\omega_i$ is decomposed in a similar way, with $\omega_{\pm 1}$ 
the corresponding components 
(in this case in the Poisson gauge there is no scalar component). 
Similarly for the tensor perturbation $\chi_{ij}$ 
we have indicated its amplitudes as $\chi_{\pm 2}$ in the 
decomposition~\cite{huwhite}
\begin{equation}
\chi_{ij}=\sum_{m=\pm 2} - \sqrt{\frac{3}{8}} \, \chi_m ({\bf e}_1\pm i 
{\bf e}_2)_i ({\bf e}_1\pm i {\bf e}_2)_j\, .  
\end{equation}
We have taken into account that in the gravitational part of the Boltzmann 
equation and in the collsion term there are some terms, like 
$\delta^{(1)}_e {\bf v}$, which still can be decomposed in the scalar and 
transverse parts in Fourier space as in Eq.~(\ref{vdec}). 
For a generic quantity  $f({\bf x}) {\bf v}$ we have indicated the 
corresponding scalar and vortical components with $(f {\bf v})_m$ and 
their explicit expression is easily found by projecting the Fourier 
modes of $f({\bf x}) {\bf v}$ along the ${\bf {\hat k}}={\bf e}_3$ and 
$({\bf e}_2\mp i {\bf e}_1)$ directions 
\begin{equation}
(f {\bf v})_m({\bf k})=\int \frac{d^3k_1}{(2\pi)^3} v^{(1)}_0({\bf k}_1) 
f({\bf k}_2) Y^*_{1m}({\bf \hat{k}}_1) \sqrt{\frac{4\pi}{3}}\, .
\end{equation} 
Similarly for a term like $f({\bf x}) \nabla g({\bf x})$ we used the notation
\begin{equation}
(f\nabla g)_m({\bf k})=- \int \frac{d^3k_1}{(2\pi)^3} k_1 g({\bf k}_1) 
f({\bf k}_2) Y^*_{1m}({\bf \hat{k}}_1) 
\sqrt{\frac{4\pi}{3}}\, .
\end{equation}   
Finally, the first term on the right-hand side of Eq.~(\ref{H}) has been 
obtained by using the relation 
\begin{eqnarray}
\label{gradient}
i {\bf k} \cdot {\bf n}\, \Delta^{(2)}({\bf k}) &=& \sum_{\ell m} 
\Delta^{(2)}_{\ell m}({\bf k}) \frac{k}{2\ell +1} \left[ 
{\kappa_{\ell m} {\tilde G}_{\ell-1,m}-\kappa_{\ell+1,m}{\tilde G}_{\ell+1,m}} 
\right] \nonumber \\
&=&  
k \sum_{\ell m} \left[\frac{{\kappa}_{\ell m}}{2\ell-1}
\Delta^{(2)}_{\ell-1,m}-
\frac{{\kappa}_{\ell m}}{2\ell+3}\Delta^{(2)}_{\ell+1,m} 
\right] {\tilde G}_{\ell m}\, , 
\end{eqnarray} 
where $ {\tilde G}_{\ell m}= (-i)^{\ell} \sqrt{4\pi/(2\ell+1)}
Y_{\ell m}({\bf n})$ is the 
angular mode for the decomposition~(\ref{Dlm2}) and 
${\kappa}_{\ell m}=\sqrt{l^2-m^2}$.
This relation has been discussed in Refs.~\cite{Complete,huwhite} 
and corresponds to the term $n^i \partial \Delta^{(2)}/\partial x^i$ 
in Eq.~(\ref{B2}). 

As expected, at second order we recover some intrinsic effects which 
are characteristic of the linear regime. 
In Eq.~(\ref{H}) the relation~(\ref{gradient})  represents the free 
streaming effect: when the radiation undergoes free-streaming, 
the inhomogeneities of the photon    
distribution are seen by the obsever as angular anisotropies. At first 
order it is responsible for the hierarchy of Boltzmann 
equations coupling the different $\ell$ modes, and it represents a 
projection effect of fluctuations on a scale $k$ onto the angular scale 
$\ell \sim k\eta$. The term $\tau' \Delta^{(2)}_{\ell m}$ causes an 
exponential suppression of anisotropies in the absence of the source 
term $S_{\ell m}$. The first line of the source term~(\ref{Slm}) just 
reproduces the expression of the first order case.   
Of course the dynamics of the second-order metric and baryon-velocity 
perturbations which appear will be different and governed by the 
second-order Einstein equations and continuity equations. The remaining 
terms in the source are second-order effects generated as 
non-linear combinations of the primordial (first-order) perturbations. We 
have ordered them according to the increasing number of 
$\ell$ modes they contribute to. Notice in particular that they involve 
the first-order anisotropies $\Delta^{(1)}_\ell$  and as a 
consequence such terms contribute to generate the hierarchy of equations 
(apart from the free-streaming effect). The source term 
contains additional scattering processes and gravitational effects. 
On large scales (above the horizon at recombination) 
we can say that the main effects are due to gravity, and they include 
the Sachs-Wolfe and the (late and early) Sachs-Wolfe effect due 
to the redshift photons suffer when travelling through the second-order 
gravitational potentials. These, toghether with the contribution 
due to the second-order tensor modes, have been already studied in 
details in Ref.~\cite{fulllarge}. 
Another important gravitational effect is 
that of lensing of photons as they travel from the last scattering 
surface to us. A contribution of this type is given by the last term
of Eq.~(\ref{Slm}). 

%%%%%%%%%%%%%%%%%%%%%%%%%%%%%%%%%%%%%%%%%%%%%%%%%%%%%%%%%%%%%%%%%%%%%%%
\subsection{Integral solution of the second-order Boltzmann equation}
%%%%%%%%%%%%%%%%%%%%%%%%%%%%%%%%%%%%%%%%%%%%%%%%%%%%%%%%%%%%%%%%%%%%%%%%%%%%
 
As in linear theory, one can derive an integral solution of the 
Boltzmann equation~(\ref{B2}) in terms of the source term $S$.  
Following the standard procedure (see {\it e.g.} Ref.~\cite{Dodelsonbook,SeZ}) 
for linear perturbations, 
we write the left-hand side as $\Delta^{(2) \prime}+ik\mu \Delta^{(2)}-\tau' 
\Delta^{(2)}=
e^{-ik\mu \eta + \tau} d[\Delta^{(2)} e^{ik \mu \eta-\tau}]/d\eta$ in order to 
derive the integral solution  
\begin{eqnarray}
\label{IS}
\Delta^{(2)}({\bf k},{\bf n},\eta_0)=\int_0^{\eta_0} d\eta 
S({\bf k},{\bf n},\eta) e^{ik\mu(\eta-\eta_0)-\tau}\, ,
\end{eqnarray}
where $\eta_0$ stands by the present time. The expression of the 
photon moments $\Delta^{(2)}_{\ell m}$ can be 
obtained as usual from Eq.~(\ref{angular1}). In the previous section 
we have already found the coefficients for the 
decomposition of source term $S$  
\begin{eqnarray}
\label{decS}
S({\bf k},{\bf n}, \eta)= \sum_{\ell} \sum_{m=-\ell}^{\ell} 
S_{\ell m}({\bf k},\eta) 
(-i)^{\ell}\ \sqrt{\frac{4\pi}{2\ell+1}} Y_{\ell m}({\bf n}) \, .
\end{eqnarray}
In Eq.~(\ref{IS}) there is an additional angular dependence in 
the exponential. It is easy to 
take it into account by recalling that 
\begin{equation}
\label{eikx}
e^{i{\bf k} \cdot {\bf x}}=
\sum_\ell (i)^\ell (2\ell+1) j_\ell(kx) P_\ell( {\bf {\hat k}} \cdot 
{\bf {\hat x}}) \, .
\end{equation}
Thus the angular integral~(\ref{angular1}) is computed by using the 
decomposition of the source term~(\ref{decS}) 
and Eq.~(\ref{eikx}) 
\begin{eqnarray}
\label{IS1}
\Delta^{(2)}_{\ell m}( {\bf k},\eta_0)&=&(-1)^{-m}(-i)^{-\ell} (2\ell+1) 
\int_0^{\eta_0} d\eta \, e^{-\tau(\eta)} \nonumber \\
&\times& 
\sum_{\ell_2}\sum_{m_2=-\ell_2}^{\ell_2} (-i)^{\ell_2} S_{\ell_2 m_2}  
\sum_{\ell_1} i^{\ell_1} j_{\ell_1}(k(\eta-\eta_0))
\nonumber \\  
&\times& (2\ell_1+1) 
\left(\begin{array}{ccc}\ell_1&\ell_2&\ell\\0&0&0\end{array}\right)
\left(\begin{array}{ccc}\ell_1&\ell_2&\ell\\0&m_2&- m\end{array}\right) \, ,
\end{eqnarray}
where the Wigner $3-j$ symbols appear because of the Gaunt integrals
\begin{eqnarray}
  \nonumber
  {\mathcal G}_{l_1l_2l_3}^{m_1m_2m_3}
  &\equiv&
  \int d^2\hat{\mathbf n}
  Y_{l_1m_1}(\hat{\mathbf n})
  Y_{l_2m_2}(\hat{\mathbf n})
  Y_{l_3m_3}(\hat{\mathbf n})\\
 \nonumber
  &=&\sqrt{
   \frac{\left(2l_1+1\right)\left(2l_2+1\right)\left(2l_3+1\right)}
        {4\pi}
        }
\nonumber \\
&\times &   \left(
  \begin{array}{ccc}
  \ell_1 & \ell_2 & \ell_3 \\ 0 & 0 & 0 
  \end{array}
  \right)
  \left(
  \begin{array}{ccc}
  \ell_1 & \ell_2 & \ell_3 \\ m_1 & m_2 & m_3 
  \end{array}
  \right),
\end{eqnarray}
Since the second of the Wigner 3-$j$ symbols in Eq.~(\ref{IS1}) 
is nonzero only if $m=m_2$, our solution can be rewritten to 
recover the corresponding expression found for linear anisotropies in 
Refs.~\cite{huwhite,Complete}
\begin{equation}
\label{comp}
\frac{\Delta^{(2)}_{\ell m}( {\bf k},\eta_0)}{2\ell+1}=
\int_0^{\eta_0} d\eta \, e^{-\tau(\eta)} \sum_{\ell_2} 
S_{\ell_2 m}\, j_{\ell}^{(l_2,m)}[k(\eta_0-\eta)]\, ,
\end{equation}    
where $j_{\ell}^{(l_2,m)}[k(\eta_0-\eta)]$ are the so called radial 
functions. Of course the main information at second order is 
included in the source term containing different effects due to the 
non-linearity of the perturbations. 
In the total angular momentum method of Refs.~\cite{huwhite,Complete} 
Eq.~(\ref{comp}) is interpreted just as the intergration over 
the radial coordinate $(\chi=\eta_0-\eta)$ of the projected source term. 
Another important comment is that, as in linear theory, the integral 
solution~(\ref{IS1}) is in fact just a formal solution, since 
the source term $S$ contains itself the second-order photon moments up
 to $l=2$ (see Eq.~(\ref{Slm})). 
This means that one has anyway to resort to the 
hierarchy equations for photons, Eq.~(\ref{H}), to solve for these 
moments. Neverthless, as in linear theory~\cite{SeZ}, 
one expects to need just a few moments beyond $\ell=2$ in the hierarchy 
equations, and once the moments entering in the 
source function are computed the higher moments are obtained from the 
integral solution. Thus the integral solution should in fact be 
more advantageous than solving the system of coupled 
equations~(\ref{H}).

%%%%%%%%%%%%%%%%%%%%%%%%%%%%%%%%%%%%%%%%%%%%%%%%%%%%%%%%%%%%%%%%%%%%%%%%
\section{The Boltzmann equation for baryons and cold dark matter}
%%%%%%%%%%%%%%%%%%%%%%%%%%%%%%%%%%%%%%%%%%%%%%%%%%%%%%%%%%%%%%%%%%%%%%%%%%
 
In this section we will derive the Boltzmann equation for massive particles, 
which is the case of interest for baryons and dark matter. These equations
 are necessary to find the time evolution of number 
densities and velocities of the baryon fluid which appear in the 
brightness equation, thus allowing to close the system of 
equations. Let us start from the baryon component. 
Electrons are tightly coupled to protons via Coulomb interactions. 
This forces the relative energy density contrasts and the 
velocities to a common value, $\delta_e=\delta_p \equiv \delta_b$ and 
${\bf v}_e={\bf v}_p \equiv {\bf v}$, so that we can identify  
electrons and protons collectively as ``baryonic'' matter. 

To derive the Boltzmann equation for baryons let us first focus 
on the collisionless equation and compute therefore $dg/d\eta$, where $g$ 
is the distribution function for a massive 
species with mass $m$. One of the differences with respect to photons is 
just that baryons are non-relativistic for the epochs of 
interest. Thus the first step is to generalize the formulae in Section 4
up to Eq.~(\ref{dni}) to the case of a massive particle. 
In this case one enforces the constraint $Q^2=g_{\mu\nu}Q^\mu Q^\nu=-m^2$ 
and it also useful to use the particle energy
$E=\sqrt{q^2+m^2}$, 
where $q$ is defined as in Eq.~(\ref{defp}). Moreover in this case it is 
very convenient to take 
the distribution function as a function of the variables $q^i=qn^i$, the 
position $x^i$ and time $\eta$, without using the explicit splitting into 
the magnitude of the momentum $q$ (or the energy E) and its direction $n^i$. 
Thus the total time derivative of the distribution functions reads
\begin{equation}
\label{DG}
\frac{d g}{d \eta}=\frac{\partial g}{\partial \eta}+
\frac{\partial g}{\partial x^i} \frac{d x^i}{d \eta}+
\frac{\partial g}{\partial q^i} \frac{d q^i}{d \eta}\, .
\end{equation}
We will not give the details of the calculation since we just need to 
replicate the 
same computation we did for the photons. For the four-momentum of the 
particle notice that $Q^i$ has the same form as Eq.~(\ref{Pi}), 
while for $Q^0$ we find
\begin{equation}
\label{P0m}
Q^0=\frac{e^{-\Phi}}{a}\, E \left(1+\omega_i \frac{q^i}{E}  \right)\, .
\end{equation}  
In the following we give the expressions for $dx^i/d\eta$ and $dq^i/d\eta$.
\\
\noindent
a) As in Eq.~(\ref{dxi}) $dx^i/d\eta=Q^i/Q^0$ and it turns out to be
\begin{equation}
\frac{d x^i}{d\eta}=\frac{q}{E} n^i e^{\Phi+\Psi}  
\left(1-\omega_i n^i \frac{q}{E} \right) 
\left(1-\frac{1}{2} \chi_{km} n^k n^m \right)\, .
\end{equation}

\noindent
b) For $dq^i/d\eta$ we need the expression of $Q^i$ which is 
the same as that of  
Eq.~(\ref{Pi}) 
\begin{equation}
\label{Qi}
Q^i= \frac{q^i}{a} e^{\Psi} \left( 1-\frac{1}{2} \chi_{km}n^kn^m \right)\, .
\end{equation}
The spatial component of the geodesic equation, up to second order, reads 
\begin{eqnarray}
\label{SPG}
\frac{dQ^i}{d\eta}&=&-2({\cal H}-\Psi')\left( 1-\frac{1}{2}\chi_{km}n^kn^m 
\right) 
\frac{q}{a}n^i e^{\Psi} + e^{\Phi+2\Psi} 
\nonumber\\
&\times & \biggl( \frac{\partial \Psi}{\partial x^k}
\frac{q^2}{aE}(2 n^in^k - \delta^{ik})
-\frac{\partial \Phi}{\partial x^i}\frac{E}{a} \biggr) 
- \frac{E}{a}\bigl[\omega^{i\prime}+{\cal H} \omega^i + 
q^k \bigl(\chi^{i\prime}_{~k} \nonumber \\
&+& \omega^{i}_{'k}-\omega_k^{,i}\bigr)\bigr]  
+\left[{\cal H}\omega^i \delta_{jk}-\frac{1}{2}(\chi^{i}_{~j,k}+\chi^i_{~k,j} -
\chi_{jk}^{~~,i})\right] \frac{q^j q^k}{Ea}\, .
\end{eqnarray}
Proceeding as in the massless case we now take the total time derivative 
of Eq.~(\ref{Qi}) and using Eq.~(\ref{SPG}) we find 
\begin{eqnarray}
\frac{dq^i}{d\eta}&=&-({\cal H}-\Psi')q^i+\Psi_{,k}\frac{q^iq^k}{E} 
e^{\Phi+\Psi}-\Phi^{,i}E e^{\Phi+\Psi} 
-\Psi_{,i}\frac{q^2}{E} e^{\Phi+\Psi} \nonumber \\
&-& E(\omega^{i\prime}+{\cal H} \omega^i)
- (\chi^{i\prime}_{~k}+\omega^{i}_{'k}-\omega_k^{,i}) E q^k \nonumber \\
&+&\left[{\cal H}\omega^i \delta_{jk}- \frac{1}{2} 
(\chi^{i}_{~j,k}+\chi^i_{~k,j} - \chi_{jk}^{~,i}) \right] \frac{q^j q^k}{E}\, .
\end{eqnarray}
We can now write the total time derivative of the distribution function as
\begin{eqnarray}
\label{Dg}
\frac{d g}{d \eta}&=& 
 \frac{\partial g}{\partial \eta}+\frac{q}{E} n^i e^{\Phi+\Psi}
\left(1-\omega_i n^i -\frac{1}{2} 
\chi_{km}n^kn^m \right) \frac{\partial g}{\partial x^i} \nonumber \\
&+& \biggl[ -({\cal H}-\Psi')q^i+\Psi_{,k}\frac{q^iq^k}{E} e^{\Phi+\Psi}-
\Phi^{,i}E e^{\Phi+\Psi} 
-\Psi_{,i}\frac{q^2}{E} e^{\Phi+\Psi} \nonumber \\
&-& E(\omega^{i\prime}+{\cal H} \omega^i)
- (\chi^{i\prime}_{~k}+\omega^{i}_{'k}-\omega_k^{,i}) E q^k \nonumber \\
&+& 
\biggl({\cal H}\omega^i \delta_{jk}-\frac{1}{2}(\chi^{i}_{~j,k}+\chi^i_{~k,j} -
\chi_{jk}^{~,i})\biggr) \frac{q^j q^k}{E} 
\biggr] \frac{\partial g}{\partial q^i}\, .
\end{eqnarray}
This equation is completely general since we have just solved for the 
kinematics of massive particles. 
As far as the collision terms are concerned, for the system of electrons 
and protons we consider the Coulomb scattering processes 
between the electrons and protons and the Compton scatterings between 
photons and electrons 
\begin{eqnarray}
\label{boltzep}
\frac{dg_e}{d\eta}({\bf x},{\bf q},\eta)&=&\langle c_{ep} \rangle_{QQ'q'} 
+\langle c_{e\gamma} \rangle_{pp'q'} \\
\label{boltzep2}
\frac{dg_p}{d\eta}({\bf x},{\bf Q},\eta)&=&\langle c_{ep} \rangle_{qq'Q'}\, ,
\end{eqnarray} 
where we have adopted the same formalism of Ref.~\cite{Dodelsonbook} 
with ${\bf p}$ and ${\bf p}'$ the initial and final momenta of the 
photons, ${\bf q}$ and ${\bf q}'$ the corresponding quantities for the 
electrons and for protons ${\bf Q}$ and ${\bf Q}'$. The 
integral  over different momenta is indicated by   
\begin{equation}
\langle \cdots \rangle_{pp'q'} \equiv \int \frac{d^3p}{(2\pi)^3}\,\int 
\frac{d^3 p'}{(2\pi)^3}\, 
\int \frac{d^3q'}{(2\pi)^3} \dots \, ,
\end{equation}
and thus one can read 
$c_{e\gamma}$ as the unintegrated part of Eq.~(\ref{collisionterm0}),
 and similarly for $c_{ep}$ (with the appropriate amplitude
$|M|^2$). In Eq.~(\ref{boltzep}) Compton scatterings between protons and 
photons can be safely neglected because the amplitude of 
this process has a much smaller amplitude than Compton scatterings with
 electrons being weighted by the inverse squared mass of the 
particles. 

At this point for the photons we considered the perturbations around the 
zero-order Bose-Einstein distribution function (which are 
the unknown quantities). For the electrons (and protons) we can take the 
thermal distribution described by Eq.~(\ref{gel}). 
Moreover we will take the moments of Eqs.~(\ref{boltzep})-(\ref{boltzep2}) 
in order to find the energy-momentum continuity equations.

%%%%%%%%%%%%%%%%%%%%%%%%%%%%%%%%%%%%%%%%%%%%%%%%%%%%%%%%%%%%%%%%%%
\subsection{Energy continuity equations} 
%%%%%%%%%%%%%%%%%%%%%%%%%%%%%%%%%%%%%%%%%%%%%%%%%%%%%%%%%%%%%%%%%%
 
We now integrate Eq.~(\ref{Dg}) over $d^3q/(2\pi)^3$. Let us recall that 
in terms of the distribution function 
the number density $n_e$ and the bulk velocity ${\bf v}$ are given by
\begin{equation}
\label{defne}
n_e=\int \frac{d^3 q}{(2\pi)^3}\, g \, , 
\end{equation} 
and 
\begin{equation}
\label{defvg}
v^i= \frac{1}{n_e} \int \frac{d^3q}{(2\pi)^3}\, g \, \frac{q n^i}{E}\, ,
\end{equation}
where one can set $E\simeq m_e$ since we are considering non-relativistic 
particles. 
We will also make use of the following relations when integrating over the 
solid angle $d\Omega$
\begin{equation}
\label{relOmega}
\int d\Omega\, n^i=\int d\Omega\, n^in^jn^k=0\, ,\quad \int 
\frac{d\Omega}{4\pi}\, n^in^j =  \frac{1}{3} \delta^{ij}\, . 
\end{equation}
Finally notice that $dE/dq=q/E$ and $\partial g/\partial q= (q/E) 
\partial g/\partial E$.

Thus the first two integrals just brings $n'_e$ and $(n_e v^i)_{,i}$. 
Notice that all the terms proportional to the second-order 
vector and tensor perturbations of the metric give a vanishing contribution 
at second order since in this case we can take the 
zero-order distribution functions which depends only on $\eta$ and $E$, 
integrate over the direction 
and use the fact that $\delta^{ij} \chi_{ij}=0$. The trick to solve 
the remaining integrals is an integration  by parts over $q^i$.  
We have an integral like (the one multiplying $( \Psi'-{\cal H})$)
\begin{equation}
\label{r1}
\int \frac{d^3q}{(2 \pi)^3} q^i \frac{\partial g}{\partial q^i} = 
-3 \int \frac{d^3q}{(2 \pi)^3} g =-3 n_e \, ,
\end{equation}
after an integration by parts over $q^i$. The remaining integrals can 
be solved still by integrating by parts over $q^i$.  
The integral proportional to $\Phi^{,i}$ in Eq.~(\ref{Dg}) gives  
\begin{equation}
\int \frac{d^3q}{(2 \pi)^3} E =- v_i \, n_{e}\, ,
\end{equation}
where we have used the fact that $dE/dq^i=q^i/E$. For the integral 
\begin{equation}
\int \frac{d^3q}{(2 \pi)^3} \frac{q^iq^k}{E} 
\frac{\partial g}{\partial q^i}\, , 
\end{equation}
the integration by parts brings two pieces, one from the derivation 
of $q^iq^k$ and one from the derivation of the energy $E$
\begin{equation}
\label{exint}
- 4 \int \frac{d^3q}{(2 \pi)^3} g\frac{q^k}{E} +     
\int \frac{d^3q}{(2 \pi)^3} g \frac{q^2}{E} \frac{q^k}{E} 
=-4 v^k\, n_e + 
\int \frac{d^3q}{(2 \pi)^3} g \frac{q^2}{E^2} \frac{q^k}{E} \, .
\end{equation}
The last integral in Eq.~(\ref{exint}) can indeed be neglected. To check 
this one makes use of the explicit  expression~(\ref{gel}) for the 
distribution function $g$ to derive 
\begin{equation}
\frac{\partial g}{\partial v^i}=g\frac{q_i}{T_e}-\frac{m_e}{T_e}v_ig\, ,
\end{equation} 
and 
\begin{equation}
\label{gpp}
\int \frac{d^3q}{(2 \pi)^3} g q^i q^j =\delta^{ij} n_e m_e T_e+ n_e m_e^2 
v^i v^j\, .
\end{equation}
Thus it is easy to compute
\begin{equation}
\label{finaleener}
\frac{\Psi_{,k}}{m_e^3} \int \frac{d^3q}{(2 \pi)^3} g q^2q^k
= - \Psi_{,k}v^2\frac{T_e}{m_e} 
+3 \Psi_{,k}v_kn_e\frac{T_e}{m_e}+
\Psi_{,k}v_kv^2 \, , 
\end{equation}
which is negligible taking into account that $T_e/m_e$ is of the 
order of the thermal velocity squared.  

With these results we are now able to compute the left-hand side  of the 
Boltzmann equation~(\ref{boltzep}) integrated over 
$d^3q/(2\pi)^3$. The same operation must be done for the collision terms 
on the right hand side. For example for the 
first of the equations in~(\ref{boltzep}) this brings to the integrals 
$ \langle c_{ep} \rangle_{QQ'qq'} +\langle c_{e\gamma} \rangle_{pp'qq'}$. 
However looking at Eq.~(\ref{collisionterm}) one realizes 
that $\langle c_{e\gamma} \rangle_{pp'qq'}$ vanishes because the 
integrand is antisymmetric under the change ${\bf q} 
\leftrightarrow {\bf q'}$ and ${\bf p} 
\leftrightarrow {\bf p}'$. In fact this is simply a consequence of the 
fact that the electron number is conserved for this process. 
The same argument holds for the other term $\langle c_{ep} \rangle_{QQ'qq'}$. 
Therefore the right-hand side of 
Eq.~(\ref{boltzep}) integrated over $d^3q/(2\pi)^3$ vanishes and we can give 
the evolution equation for $n_e$. Collecting 
the results of Eq.~(\ref{r1}) to~(\ref{finaleener}) we find 
\begin{equation}
\label{cont_bar}
\frac{\partial n_e}{\partial \eta}+e^{\Phi+\Psi} 
\frac{\partial(v^i n_e)}{\partial x^i}+3(
{\cal H}-\Psi')n_e + e^{\Phi+\Psi} v^k n_e \left(\Phi_{,k} -2 \Psi_{,k}\right) 
= 0\, . 
\end{equation}
Similarly, for CDM particles, we find
\begin{eqnarray}
\frac{\partial n_{\rm CDM}}{\partial \eta}&+&e^{\Phi+\Psi} \frac{\partial(v^i 
n_{\rm CDM})}{\partial x^i}+3(
{\cal H}-\Psi')n_{\rm CDM} \nonumber \\
&+& e^{\Phi+\Psi} v_{\rm CDM}^k\,
n_{\rm CDM} \biggl( \Phi_{,k} - 2 \Psi_{,k} \biggr) = 0 \, .
\end{eqnarray}

%%%%%%%%%%%%%%%%%%%%%%%%%%%%%%%%%%%%%%%%%%%%%%%%%%%%%%%%%%%%%%%%%%%%%%%
\subsection{Momentum continuity equations}
%%%%%%%%%%%%%%%%%%%%%%%%%%%%%%%%%%%%%%%%%%%%%%%%%%%%%%%%%%%%%%%%%%%%%%%
 
Let us now multiply Eq.~(\ref{Dg}) by $(q^i/E) /(2 \pi)^3$ and 
integrate over $d^3q$. In this way we will find the continuity 
equation for the momentum of baryons. The first term just gives 
$(n_e v^i)'$. The second integral is of the type 
\begin{equation}
\frac{\partial}{\partial x^j} \int \frac{d^3q}{(2 \pi)^3} g\, 
\frac{q n^j}{E} \frac{qn^i}{E} =
\frac{\partial}{\partial x^j}\left( n_e \frac{T_e}{m_e} \delta^{ij}+n_e 
v^i v^j \right)\, ,
\end{equation} 
where we have used Eq.~(\ref{gpp}) and $E=m_e$. The third term proportional 
to $({\cal H}-\Psi')$ is 
\begin{eqnarray}
\label{FirstI}
\int \frac{d^3q}{(2 \pi)^3} q^k \frac{\partial g}{\partial q_k} \frac{q^i}{E}=
4 n_e+\int \frac{d^3q}{(2 \pi)^3} g \frac{q^2}{E^2}\frac{q^i}{E}\, ,
\end{eqnarray}
where we have integrated by parts over $q^i$. Notice that the last 
term in Eq.~(\ref{FirstI}) is negligible being the same integral we 
discussewd above in Eq.~(\ref{finaleener}). By the same arguments that 
lead to neglect the term of Eq.~(\ref{finaleener}) it is easy to 
check that all the remaining integrals proportional to the gravitational
 potentials are negligible except for 
\begin{equation}
- e^{\Phi+\Psi} \Phi_{,k}\int \frac{d^3q}{(2 \pi)^3} 
\frac{\partial g}{\partial q_k} q^i=n_ee^{\Phi+\Psi}\Phi^{,i}\, . 
\end{equation}
The integrals proportional to the second-order vector and tensor 
perturbations vanish as 
vector and tensor perturbations are traceless and divergence-free. The 
only one which survives is the term proportional to $\omega^{i\prime}
+{\cal H}\omega^{i}$ in Eq.~(\ref{Dg}). 

Therefore  for the integral over $d^3q q^i/E$ 
of the left-hand side of the Boltzmann equation~(\ref{Dg})
for a massive particle with mass $m_e$ ($m_p$) and distribution 
function~(\ref{gel}) we find 
\begin{eqnarray}   
\label{dgetau}   
&&\int \frac{d^3q}{(2\pi)^3} \frac{q^i}{E} 
\frac{d g_e}{d\eta} =  \frac{\partial (n_e v^i)}{\partial \eta}+
4 ({\cal H}-\Psi') n_e v^i +\Phi^{,i} e^{\Phi+\Psi} 
n_e \nonumber \\
&&+ e^{\Phi+\Psi} \left( n_e \frac{T_e}{m_e} \right)^{,i}+
 e^{\Phi+\Psi} \frac{\partial}{\partial x^j}(n_e v^j v^i) +
\frac{\partial \omega^i}{\partial \eta} n_e + {\cal H} \omega^i n_e \, . 
\nonumber \\
\end{eqnarray}

Now, in order to derive the momentum conservation equation for baryons, we 
take the first moment of both Eq.~(\ref{boltzep}) 
and~(\ref{boltzep2}) multiplying them by ${\bf q}$ and ${\bf Q}$ 
respectively and integrating over the momenta. Since previously we 
integrated the left-hand side of these equations over $d^3q q^i/E$, 
we just need to multiply 
the previous integrals by $m_e$ for the electrons and for $m_p$ for 
the protons. Therefore if we sum the 
first moment of Eqs.~(\ref{boltzep}) and~(\ref{boltzep2}) the dominant 
contribution on the left-hand side will be that of the protons
\begin{eqnarray}
\int \frac{d^3 Q}{(2 \pi)^3} Q^i\, \frac{dg_p}{d\eta}=\langle c_{ep} 
(q^i+Q^i)\rangle_{QQ'qq'} +\langle c_{e\gamma} q^i\rangle_{pp'qq'}\, .
\end{eqnarray}  
Notice that the integral of the Coulomb collision term $c_{ep} (q^i+Q^i)$ 
over all momenta vanishes simply because of momentum conservation 
(due to the Dirac function $\delta^4(q+Q-q'-Q')$). As far as the Compton 
scattering is concerned we have that, following 
Ref.~\cite{Dodelsonbook}, 
\begin{equation}
\langle c_{e\gamma} q^i \rangle_{pp'qq'} =- \langle c_{e\gamma} p^i 
\rangle_{pp'qq'} \, ,  
\end{equation}   
still because of the total momentum conservation. Therefore what we 
can compute now is the integral over all momenta of 
$c_{e\gamma} p^i$. Notice however that this is equivalent just to 
multiply the Compton collision term $C({\bf p})$ 
of Eq.~(\ref{collisionterm}) by $p^i$ and integrate over $d^3p/(2\pi^3)$ 
\begin{equation}
\label{Ci}
\langle c_{e\gamma} p^i \rangle_{pp'qq'} = a e^{\Phi} \int 
\frac{d^3p}{(2\pi)^3} p^i C({\bf p})\, .
\end{equation}
where $C({\bf p})$ has been already computed in Eqs.~(\ref{C1p}) 
and~(\ref{C2p}). 

We will do the integral~(\ref{Ci}) in the following. First let us 
introduce the definition of the velocity of photons in terms of 
the distribution function 
\begin{equation}
\label{vp}
(\rho_\gamma+p_\gamma) v^i_\gamma = \int \frac{d^3 p}{(2 \pi)^3} f p^i\, ,
\end{equation}  
where $p_\gamma = \rho_\gamma/3$ is the photon pressure and 
$\rho_\gamma$ the energy density. At first order we get
\begin{equation}
\label{vp1}
\frac{4}{3} v^{(1) i}_\gamma= \int \frac{d\Omega}{4\pi} 
\Delta^{(1)}\, n^i \, ,
\end{equation}
where $\Delta$ is the photon distribution anisotropies defined 
in Eq.~(\ref{Delta2}). At second order we instead find
\begin{equation}
\label{vp2}
\frac{4}{3} \frac{v^{(2) i}_\gamma}{2}= \frac{1}{2} \int 
\frac{d\Omega}{4\pi} \Delta^{(2)}\, n^i-\frac{4}{3} \delta^{(1)}_\gamma 
v^{(1)i}_\gamma \, .
\end{equation}
Therefore the terms in Eqs.~(\ref{C1p}) and~(\ref{C2p}) 
proportional to $f^{(1)}({\bf p})$ and $f^{(2)}({\bf p})$ 
will give rise to terms containing the velocity of the photons. 
On the other hand 
the terms proportional to $f^{(1)}_0(p)$ and $f^{(2)}_{00}(p)$, 
once integrated, vanish because of the integral over the momentum direction
$n^i$, $\int d\Omega n^i=0$. Also the integrals involving 
$P_2({\bf {\hat v}}\cdot {\bf n})=[3 ({\bf {\hat v}}\cdot 
{\bf n})^2-1]/2$ 
in the first line of Eq.~(\ref{C1p}) and~(\ref{C2p}) 
vanish since 
\begin{equation}
\int d\Omega P_2({\bf {\hat v}}\cdot {\bf n})\,  n^i= {\hat v}^k 
{\hat v}^j  \int d\Omega n_kn_jn^i=0\, ,
\end{equation}
where we are using the relations~(\ref{relOmega}). Similarly all 
the terms proportional to $v$, 
$({\bf v} \cdot {\bf n})^2$ and $v^2$ do not give any contribution 
to Eq.~(\ref{Ci}) and, 
in the second-order collision term, one can check that 
$\int d\Omega Y_2({\bf n}) n^i =0$. Then there are terms proportional to 
$({\bf v}\cdot {\bf n}) f^{(0)}(p)$, $({\bf v}\cdot {\bf n}) p 
\partial f^{(0)}/\partial p$ and 
$({\bf v}\cdot {\bf n}) p \partial f^{(1)}_0/\partial p$ for which 
we can use the rules~(\ref{rules}) when 
integrating over $p$ while the integration over the momentum
 direction is   
\begin{equation}
\int \frac{d\Omega}{4\pi} ({\bf v}\cdot {\bf n}) n^i =v_k  \int 
\frac{d\Omega}{4\pi}  n^kn^i= \frac{1}{3} v^i \, .
\end{equation}
Finally from the second line of Eq.~(\ref{C2p}) we get three integrals. One is 
\begin{equation}
\int \frac{d^3p}{(2\pi)^3} p^i \, ({\bf v}\cdot {\bf n}) 
f^{(1)}({\bf p})= \bar{\rho}_\gamma 
\int \frac{d\Omega}{4 \pi} \Delta^{(1)} ({\bf v}\cdot {\bf n})  n^i\, , 
\end{equation} 
where $\bar{\rho}_\gamma$ is the background energy density of the photons. 
The second comes from  
\begin{eqnarray}
&&\frac{1}{2} \int \frac{d^3p}{(2\pi)^3} p^i\,   ({\bf v} 
\cdot {\bf n}) P_2({\bf {\hat v}}\cdot {\bf n}) \left(f^{(1)}_2(p)-
p \frac{\partial f^{(1)}_2(p) }{\partial p} \right) \\
&&= \frac{5}{4} 
\bar{\rho}_\gamma \Delta^{(1)}_2 
\left[ 3 
v_j  {\hat v}_k {\hat v}_l \int \frac{d\Omega}{4 \pi} n^in^jn^kn^l-v_j 
\int \frac{d\Omega}{4 \pi} n^in^j 
\right] = \frac{1}{3} \bar{\rho}_\gamma \Delta^{(1)}_2 {\hat v}^i\, ,
\nonumber 
\end{eqnarray}
where we have used the rules~(\ref{rules}), Eq.~(\ref{relOmega}) and 
$\int (d\Omega/4 \pi)\, n^in^jn^kn^l = (\delta^{ij} \delta^{kl} 
+\delta^{ik} \delta^{lj}+\delta^{il} \delta^{jk})/15$. In fact the 
third integral 
\begin{equation}
- \int \frac{d^3p}{(2\pi)^3} p^i ({\bf v} \cdot {\bf n}) f^{(1)}_2(p)\, , 
\end{equation}
exactly cancels the previous one. Summing the various integrals we find  
\begin{eqnarray}
\label{Ciint}
&&\int\frac{d{\bf p}}{(2\pi)^3} C({\bf p}) {\bf p}= 
n_e \sigma_T {\bar \rho_\gamma} \Bigg[ \frac{4}{3} 
({\bf v}^{(1)}-{\bf v}_\gamma^{(1)})
-\int \frac{d\Omega}{4\pi} \frac{\Delta^{(2)}}{2} {\bf n} \nonumber\\
&&+ \frac{4}{3} 
\frac{{\bf v}^{(2)}}{2} 
+\frac{4}{3} \delta^{(1)}_e ({\bf v}^{(1)}-{\bf v}_\gamma^{(1)})+ 
\int \frac{d\Omega}{4\pi} \Delta^{(1)} ({\bf v} \cdot {\bf n}) {\bf n} 
+ \Delta_0^{(1)} {\bf v} \Bigg] \;.
\end{eqnarray}
Eq.~(\ref{Ciint}) can be further simplified. Recalling that 
$\delta^{(1)}_\gamma = \Delta^{(1)}_0$ we use Eq.~(\ref{vp2}) and notice that 
\begin{equation}
\int \frac{d\Omega}{4\pi} \Delta^{(1)}\, ({\bf v} \cdot {\bf n}) 
n^i = v_j^{(1)} \Pi^{ji}_\gamma+\frac{1}{3} v^i \Delta^{(1)}_0 \, ,
\end{equation}
where the photon quadrupole $\Pi^{ij}_\gamma$ is defined as 
\begin{equation}
\label{quadrupole}
\Pi^{ij}_{\gamma}=\int\frac{d\Omega}{4\pi}\,\left(n^i n^j-\frac{1}{3}
\delta^{ij}\right)\left(\Delta^{(1)}+\frac{\Delta^{(2)}}{2}\right)\, . 
\end{equation}

Thus, our final expression for the integrated collision term~(\ref{Ci}) reads
\begin{eqnarray}
\label{Cifinal}
&&\int \frac{d^3 p}{(2\pi)^3} C({\bf p}) p^i =
n_e \sigma_T {\bar \rho_\gamma}
\left[ \frac{4}{3} (v^{(1)i}-v_\gamma^{(1)i}) + \frac{4}{3} 
\left( \frac{v^{(2)i}}{2}-\frac{v_\gamma^{(2)i}}{2} \right) 
\right. \nonumber \\ && 
+ \left. \frac{4}{3} \left(\delta^{(1)}_e +\Delta^{(1)}_0\right) (v^{(1)i}-
v_\gamma^{(1)i})+ v^{(1)}_j \Pi^{ji}_\gamma \right ]\, .
\end{eqnarray}

We are now able to give the momentum continuity equation for baryons
by combining $m_p dg_p/d\eta$ from Eq.~(\ref{dgetau}) with the 
collision term~(\ref{Ci})
\begin{eqnarray}
\label{mcc}
\frac{\partial (\rho_b v^i)}{\partial \eta} &+&4 ({\cal H}-\Psi') 
\rho_b v^i +\Phi^{,i} e^{\Phi+\Psi} 
\rho_b+ e^{\Phi+\Psi} \left( \rho_b \frac{T_b}{m_p}\right)^{,i}
\nonumber \\
& + & e^{\Phi+\Psi} \frac{\partial}{\partial x^j}(\rho_b v^j v^i) +
\frac{\partial \omega^i}{\partial \eta} \rho_b+ {\cal H} \omega^i 
\rho_b  \nonumber \\
&= & - n_e \sigma_T a\, {\bar \rho_\gamma}
\left[ \frac{4}{3} (v^{(1)i}-v_\gamma^{(1)i}) + \frac{4}{3} 
\left( \frac{v^{(2)i}}{2}-\frac{v_\gamma^{(2)i}}{2} \right) 
\right. \nonumber \\
&+& \left. 
\frac{4}{3} \left( \delta^{(1)}_b+\Delta^{(1)}_0+\Phi^{(1)} 
\right) (v^{(1)i}-v_\gamma^{(1)i})+v^{(1)}_j \Pi^{ji}_\gamma \right ] \, ,
\end{eqnarray}
where $\rho_b$ is the baryon energy density and, as we previously explained, 
we took into account that to a good 
approximation the electrons do not contribute to the mass of baryons. 
In the following we will expand explicitly at first and second-order 
Eq.~(\ref{mcc}).

\subsubsection{First-order momentum continuity equation for baryons}

At first order we find
\begin{equation}
\label{mcc1}
\frac{\partial v^{(1)i}}{\partial \eta} +{\cal H} v^{(1)i}+\Phi^{(1),i}=
\frac{4}{3} \tau' \frac{{\bar \rho}_\gamma}{{\bar \rho}_b} 
\left(  v^{(1)i}-v^{(1)i}_\gamma \right)\, .
\end{equation} 

\subsubsection{Second-order momentum continuity equation for baryons}

At second order there are various simplifications. In particular notice 
that the term on the right-hand side of Eq.~(\ref{mcc}) which is proportional 
to $\delta_b$ vanishes when matched to expansion of the left-hand side by 
virtue of the first-order equation~(\ref{mcc1}). Thus, at the end 
we find a very simple equation
\begin{eqnarray}
& & \frac{1}{2} \left( \frac{\partial v^{(2)i}}{\partial \eta} 
+{\cal H} v^{(2)i} + 
2 \frac{\partial \omega^i}{\partial \eta} +2 {\cal H} \omega_i  + 
\Phi^{(2),i}\right) - \frac{\partial \Psi^{(1)}}{\partial \eta} 
v^{(1)i} \nonumber \\
&& + v^{(1)j}\partial_jv^{(1)i}+(\Phi^{(1)}+\Psi^{(1)}) \Phi^{(1),i} 
+\left( \frac{T_b}{m_p} \right)^{,i} = \frac{4}{3} \tau' 
\frac{{\bar \rho}_\gamma}{{\bar \rho}_b} \\
&& \times 
\left[ \left(  \frac{v^{(2)i}}{2}-\frac{v^{(2)i}_\gamma}{2}  \right) +
\left(\Delta^{(1)}_0+\Phi^{(1)}  \right) 
\left(  v^{(1)i}-v^{(1)i}_\gamma  \right)
+\frac{3}{4} v^{(1)}_j \Pi^{ji}_\gamma 
\right] \, \nonumber ,
\end{eqnarray}   
with $\tau'=- {\bar n}_e \sigma_T a$.

\subsubsection{First-order momentum continuity equation for CDM}

Since CDM particles are collisionless, at first order we find
\begin{equation}
\label{mcc1CDM}
\frac{\partial v_{\rm CDM}^{(1)i}}{\partial \eta} +{\cal H} 
v_{\rm CDM}^{(1)i}+\Phi^{(1),i}=0\, .
\end{equation}

\subsubsection{Second-order momentum continuity equation for CDM}

At second order we find 
\begin{eqnarray}
\label{mcc2CDM}
&&\frac{1}{2} \left( \frac{\partial v_{\rm CDM}^{(2)i}}{\partial \eta} 
+{\cal H} v_{\rm CDM}^{(2)i} + 
2 \frac{\partial \omega^i}{\partial \eta} +2 {\cal H} \omega_i  + 
\Phi^{(2),i}\right) - \frac{\partial \Psi^{(1)}}{\partial \eta} 
v_{\rm CDM}^{(1)i} \nonumber \\
&&+ v_{\rm CDM}^{(1)j}\,  \partial_j v_{\rm CDM}^{(1)i}+
(\Phi^{(1)}+\Psi^{(1)}) \Phi^{(1),i} 
+\left( \frac{T_{\rm CDM}}{m_{\rm CDM}} \right)^{,i} 
=0\, .
\end{eqnarray}   

%%%%%%%%%%%%%%%%%%%%%%%%%%%%%%%%%%%%%%%%%%%%%%%%%%%%%%%%%%%%%%%%%%%%%%%%%%%%%
\section{Linear solution of the Boltzmann equations}      
%%%%%%%%%%%%%%%%%%%%%%%%%%%%%%%%%%%%%%%%%%%%%%%%%%%%%%%%%%%%%%%%%%%%%%%%%%%%
\label{LH2LBE}

In this section we will solve the Boltzmann 
equations at first order in perturbation theory. The interested reader 
will find the extension of these formulae to second order in 
Ref.~\cite{paperII}.  
The first two moments of the photon Boltzmann equation 
are obtained by integrating Eq.~(\ref{B1}) 
over $d\Omega_{\bf n}/4\pi$ and 
$d\Omega_{\bf n} n^i /4\pi$ respectively and they lead to 
the density and velocity continuity equations 
\begin{equation}
\label{LH2B1l}
\Delta^{(1)'}_{00}+\frac{4}{3} \partial_i 
v^{(1)i}_\gamma-4\Psi^{(1)'}=0\, ,
\end{equation}
\begin{equation}
\label{LH2B2l}
v^{(1)i\prime}_\gamma+\frac{3}{4} \partial_j 
\Pi^{(1)ji}_\gamma+\frac{1}{4} \Delta^{(1),i}_{00}
+\Phi^{(1),i}=-\tau' \left( v^{(1)i}-v^{(1)}_\gamma \right)\, , 
\end{equation}
where $\Pi^{ij}$ is the photon quadrupole moment, defined in 
Eq.~(\ref{quadrupole}).

Let us recall here that $\delta^{(1)}_\gamma=\Delta^{(1)}_{00}= 
\int d\Omega \Delta^{(1)}/4\pi$ and that 
the photon velocity is given by Eq.~(\ref{vp1}).

The two equations above are complemented by the momentum 
continuity equation for baryons, which can be conveniently written 
as  
\begin{equation}
\label{LH2bv1}
v^{(1)i}=v^{(1)i}_\gamma+\frac{R}{\tau'} \left[v^{(1)i\prime}+
{\cal H} v^{(1)i} +\Phi^{(1),i}   \right]\, ,
\end{equation}
where we have introduced the baryon-photon ratio 
$R \equiv 3 \rho_b/(4\rho_\gamma)$.

Eq.~(\ref{LH2bv1}) is in a form ready for a consistent expansion 
in the small quantity $\tau^{-1}$ which can be performed 
in the tight-coupling limit. By first taking $v^{(1)i}=v^{(1)i}_\gamma$ 
at zero order and then using this relation in the 
left-hand side of Eq.~(\ref{LH2bv1}) one obtains
\begin{equation}
\label{LH2vv}
v^{(1)i}-v^{(1)i}_\gamma=\frac{R}{\tau'} \left[v^{(1)i\prime}_\gamma
+{\cal H} v^{(1)i}_\gamma +\Phi^{(1),i}   \right]\, .
\end{equation}
Such an expression for the difference of velocities can be used in 
Eq.~(\ref{LH2B2l}) to give the evolution equation for 
the photon velocity in the limit of tight coupling
\begin{equation}
\label{LH2vphotontight}
v^{(1)i\prime}_\gamma+{\cal H}\frac{R}{1+R}v^{(1)i}_\gamma 
+\frac{1}{4} \frac{\Delta^{(1),i}_{00}}{1+R}+\Phi^{(1),i} =0\, .
\end{equation}
Notice that in Eq.~(\ref{LH2vphotontight}) 
we are neglecting the quadrupole of the photon distribution $\Pi^{(1) ij}$ 
(and all the higher moments) since it is well known that at linear 
order such moment(s) are suppressed in the tight-coupling limit 
by (successive powers of) $1/\tau$ with respect to the first two 
moments, the photon energy density and velocity. 
Eqs.~(\ref{LH2B1l}) and (\ref{LH2vphotontight}) are the master equations 
which govern the photon-baryon fluid acoustic 
oscillations before the epoch of recombination when photons and baryons 
are tightly coupled by Compton scattering. 

In fact one can combine these two equations to get a single second-order 
differential equation for the photon energy 
density perturbations $\Delta^{(1)}_{00}$. 
Deriving Eq.~(\ref{LH2B1l}) with respect to conformal time and using
Eq.~(\ref{LH2vphotontight}) to replace 
$\partial_i v^{(1)i}_\gamma$ yields
\begin{eqnarray}
\label{LH2eqoscill}
&&\left( \Delta^{(1)\prime \prime}_{00}-4\Psi^{(1)\prime \prime} \right) 
+{\cal H}\frac{R}{1+R} 
\left( \Delta^{(1)\prime}_{00}-4\Psi^{(1)\prime} \right) \nonumber \\
&&-c_s^2 \nabla^2 
\left( \Delta^{(1)}_{00}-4\Psi^{(1)} \right) 
= \frac{4}{3} \nabla^2 
\left( \Phi^{(1)}+\frac{\Psi^{(1)}}{1+R} \right)\, ,
\end{eqnarray}     
where $c_s=1/\sqrt{3(1+R)}$ is the speed of sound of the photon-baryon 
fluid. Indeed, in order to solve Eq.~(\ref{LH2eqoscill}) one needs to know 
the evolution of the gravitational potentials. 
We will come back later to the discussion of the solution of 
Eq.~(\ref{LH2eqoscill}). 

A useful relation we will use in the following is obtained by 
considering the continuity equation for the 
baryon density perturbation. By perturbing at first order 
Eq.~(\ref{cont_bar}) we obtain
\begin{equation}
\label{LH2bcont}
\delta^{(1)\prime}_b+v^i_{,i}-3\Psi^{(1)\prime}=0\, .
\end{equation}
Subtracting Eq.~(\ref{LH2bcont}) form Eq.~(\ref{LH2B1l}) brings 
\begin{equation}
\Delta^{(1)\prime}_{00}-\frac{4}{3} \delta^{(1)\prime}_b+\frac{4}{3} 
(v^{(1)i}_\gamma-v^{(1)i})_{,i}=0\, ,
\end{equation} 
which implies that at lowest order in the tight-coupling approximation 
\begin{equation}
\label{LH2D100d1b}
\Delta^{(1)}_{00}=\frac{4}{3}\delta^{(1)}_b \, ,
\end{equation}
for adiabatic perturbations.
%%%%%%%%%%%%%%%%%%%%%%%%%%%%%%%%%%%%%%%%%%%%%%%%%%%%%%%%%%%%%%%%%%%%%%
\subsection{Linear solutions in the limit of tight coupling}
%%%%%%%%%%%%%%%%%%%%%%%%%%%%%%%%%%%%%%%%%%%%%%%%%%%%%%%%%%%%%%%%%%%%%
\label{LH2Tsol1}
In this section we briefly recall how to obtain at linear order the 
solutions of the Boltzmann equations~(\ref{LH2eqoscill}).  
These correspond to the acoustic oscillations of the photon-baryon fluid 
for modes which are within 
the horizon at the time of recombination.   
It is well known that, in the variable $(\Delta^{(1)}_{00}-4\Psi^{(1)})$, 
the solution can be written as~\cite{Huthesis,Husug}
\begin{eqnarray}
\label{LH2soltot}
&&[1+R(\eta)]^{1/4} (\Delta^{(1)}_{00}-4\Psi^{(1)})= 
A\, \cos[kr_s(\eta)]+B\,\sin[kr_s(\eta)] \nonumber \\
&&-4\frac{k}{\sqrt{3}} \int_0^\eta d\eta' [1+R(\eta')]^{3/4} 
\left(\Phi^{(1)}(\eta')+\frac{\Psi^{(1)}(\eta')}{1+R(\eta')} 
\right) \nonumber \\
&&\times \sin[k(r_s(\eta)-r_s(\eta'))] \; ,
\end{eqnarray}
where the sound horizon is given by
$r_s(\eta)=\int_0^\eta d\eta' c_s(\eta')$,
with $R= 3 \rho_b/(4\rho_\gamma)$. 
The constants $A$ and 
$B$ in Eq.~(\ref{LH2soltot}) are fixed by the choice of initial conditions. 

In order to give an analytical solution that catches most of the physics 
underlying Eq.~(\ref{LH2soltot}) and which remains 
at the same time very simple to treat, we will make some simplifications 
following Ref.~\cite{HZ,Dodelsonbook}. 
First, for simplicity, we are going to neglect the ratio $R$ wherever 
it appears, 
%(corresponding to a low baryon content at the time of recombination) 
{\it except} in the arguments of the varying cosines and sines, where
 we will treat $R=R_*$ as a constant 
evaluated at the time of recombination. In this way we  
keep track of a damping of the photon velocity amplitude with respect 
to the case 
$R=0$ which prevents the acoustic peaks in the power-spectrum to disappear.   
Treating $R$ as a constant is justified by the fact that for modes 
within the horizon the 
time scale of the ocillations is much shorter than the time scale 
on which $R$ varies. 
If $R$ is a constant the sound speed is just a constant 
$c_s=1/\sqrt{3(1+R_*)}$, 
and the sound horizon is simply $r_s(\eta)=c_s \eta$. 

Second, we are going to solve for the evolutions of the perturbations 
in two well distinguished 
limiting regimes. One regime is for those perturbations which enter 
the Hubble radius when matter is the dominant 
component, that is at times much bigger than the equality epoch, with 
$k \ll k_{eq} \sim \eta^{-1}_{eq}$, 
where $k_{eq}$ is the wavenumber of the Hubble radius at the equality 
epoch. The other regime is for those perturbations 
with much smaller wavelenghts which enter the Hubble radius when the 
universe is still radiation dominated, that is 
perturbations with wavenumbers $k \gg k_{eq}\sim \eta_{eq}^{-1}$. In 
fact we are interested in  
perturbation modes which are within the horizon by the time of 
recombination $\eta_*$. Therefore 
we will further suppose that $\eta_* \gg \eta_{eq}$ in order to study 
such modes in the first regime. Even though $\eta_* \gg \eta_{eq}$ is 
not the real case, it allows to obtain some  
analytical expressions. 

Before solving for these two regimes let us fix our initial conditions,  
which are taken on large scales deep in the radiation dominated era 
(for $\eta \rightarrow 0$). 
During this epoch, for adiabatic perturbations, the gravitational 
potentials remain constant on large scales 
(we are neglecting anisotropic stresses so that $\Phi^{(1)} \simeq 
\Psi^{(1)}$) and from the $(0-0)$-component of Einstein 
equations 
\begin{equation}
\label{LH2initcond1}
\Phi^{(1)}(0)=-\frac{1}{2} \Delta^{(1)}_{00}(0)\, .
\end{equation}
On the other hand, from the energy continuity equation~(\ref{LH2B1l}) 
on large scales 
\begin{equation}
\label{LH2initcond2}
\Delta^{(1)}_{00}-4\Psi^{(1)}={\rm const.}\, ;
\end{equation}
from Eq.~(\ref{LH2initcond1}) the constant on the right-hand side 
of Eq.~(\ref{LH2initcond2}) is 
fixed to be $-6 \Phi^{(1)}(0)$; thus we find $B=0$ and $A=-6 \Phi^{(1)}(0)$. 

With our simplifications Eq.~(\ref{LH2soltot}) then reads 
\begin{eqnarray}
\label{LH2solsempl}
\Delta^{(1)}_{00}-4\Psi^{(1)}=& 
-&6\Phi^{(1)}(0) \cos(\omega_0 \eta) \nonumber \\
&-& \frac{8k}{\sqrt{3}}\int_0^\eta d\eta' \Phi^{(1)}(\eta') 
\sin[\omega_0 (\eta-\eta')]\, ,
\end{eqnarray}
where $\omega_0=kc_s$. 

%%%%%%%%%%%%%%%%%%%%%%%%%%%%%%%%%%%%%%%%%%%%%%%%%%%%%%%%%%%%%%%%%%%%%
\subsection{Perturbation modes with $k \ll k_{eq}$}
%%%%%%%%%%%%%%%%%%%%%%%%%%%%%%%%%%%%%%%%%%%%%%%%%%%%%%%%%%%%%%%%%%%%
This regime corresponds to perturbation modes which enter the Hubble 
radius when the universe is matter dominated at times 
$\eta \gg \eta_{eq}$. During matter domination the gravitational potential 
remains constant (both on super-horizon and 
sub-horizon scales), as one can see for example from 
Eq.~(\ref{LH2PhiCDM}), and its value is fixed to $\Phi^{(1)}({\bf k}, 
\eta)=\frac{9}{10} \Phi^{(1)}(0)$, where $\Phi^{(1)}(0)$ 
corresponds to the gravitational potential on large scales during the 
radiation dominated epoch. Since we are interested in 
the photon anisotropies around the time of recombination, when matter 
is dominating, we can perform the integral appearing 
in Eq.~(\ref{LH2soltot}) by taking the gravitational potential equal to 
its value during matter domination so that it is 
easily computed
\begin{equation}
\label{LH2DPhimatter}
2 \int_0^\eta d\eta' \Phi^{(1)}(\eta') 
\sin[\omega_0(\eta-\eta')]=\frac{18}{10} \frac{\Phi^{(1)}(0)}{\omega_0} 
\left( 1-\cos(\omega_0 \eta) 
\right)\, .
\end{equation} 
Thus Eq.~(\ref{LH2solsempl}) gives  
\begin{equation}
\label{LH2D001sol}
\Delta^{(1)}_{00} -4\Psi^{(1)}=\frac{6}{5} \Phi^{(1)}(0)\, 
\cos(\omega_0\eta)-\frac{36}{5} \Phi^{(1)}(0)\, .
\end{equation}
The baryon-photon fluid velocity can then be obtained as 
$\partial_i v^{(1)i}_\gamma=- 3 (\Delta^{(1)}_{00}-4\Psi^{(1)})'/4$
from Eq.~(\ref{LH2B1l}). In Fourier space 
\begin{equation}
ik_i\, v^{(1)i}_\gamma=\frac{9}{10} \Phi^{(1)}(0) \sin(\omega_0 \eta) 
\omega_0\, ,
\end{equation}  
where, going to Fourier space, $\partial_i v^{(1)i}_\gamma \rightarrow i k_i 
\, v^{(1)i}_\gamma({\bf k})$ and
\begin{equation}
\label{LH2v1sol}
v^{(1)i}_\gamma=-i \frac{k^i}{k}\frac{9}{10} \Phi^{(1)}(0) \sin(\omega_0 
\eta) c_s\, ,
\end{equation}
since the linear velocity is irrotational. 
 
%%%%%%%%%%%%%%%%%%%%%%%%%%%%%%%%%%%%%%%%%%%%%%%%%%%%%%%%%%%%%%%%%%
\subsection{Perturbation modes with $k \gg k_{eq}$}
%%%%%%%%%%%%%%%%%%%%%%%%%%%%%%%%%%%%%%%%%%%%%%%%%%%%%%%%%%%%%%%%%%
This regime corresponds to perturbation modes which enter the Hubble 
radius when the universe is still radiation dominated 
at times $\eta \ll \eta_{eq}$. In this case an approximate analytical
 solution for 
the evolution of the perturbations can be obtained by considering the 
gravitational potential for a pure 
radiation dominated epoch, given by Eq.~(\ref{LH2Phir}). For the integral 
in Eq.~(\ref{LH2solsempl}) we thus find 
\begin{equation}
\int_0^\eta \Phi^{(1)}(\eta') \sin[\omega_0(\eta-\eta')] = 
-\frac{3}{2\omega_0} \cos(\omega_0 \eta)\, , 
\end{equation}
where we have kept only the dominant contribution oscillating in 
time, while neglecting terms which decay in time. The 
solution~(\ref{LH2solsempl}) becomes
\begin{equation}
\label{LH2DPsrd}
\Delta^{(1)}_{00} -4\Psi^{(1)}=6 \Phi^{(1)}(0)\, \cos(\omega_0\eta)\, ,
\end{equation} 
and the velocity is given by
\begin{equation}
\label{LH2vsrd}
v^{(1)i}_\gamma=-i \frac{k^i}{k}\frac{9}{2} \Phi^{(1)}(0) 
\sin(\omega_0 \eta) c_s\, ,
\end{equation}

Notice that the solutions~(\ref{LH2DPsrd})--(\ref{LH2vsrd}) 
are actually correct only when radiation dominates.
Indeed, between the epoch of equality and recombination, matter 
starts to dominate. 
Full account of such a period is given e.g. in Section 7.3 of 
Ref.~\cite{Dodelsonbook}, while its consequences for the CMB anisotropy 
evolution can be found e.g. in Ref.~\cite{mukhanov}. 

%%%%%%%%%%%%%%%%%%%%%%%%%%%%%%%%%%%%%%%%%%%%%%%%%%%%%%%%%%%%%%%%%%%%%%%%%%%
\section{Conclusions}
%%%%%%%%%%%%%%%%%%%%%%%%%%%%%%%%%%%%%%%%%%%%%%%%%%%%%%%%%%%%%%%%%%%%%%%%%%

\begin{table}[ht]
\centering
\begin{tabular}{c c c c c c}
\hline\hline 
\vspace{0.2cm}
Symbol & Definition & Equation \\
\hline  
$\Phi,\Psi$ & Gravitational potentials in Poisson gauge & 
(\ref{metric}) \\
$\omega_i$ & $2$nd-order vector perturbation in Poisson gauge & 
(\ref{metric}) \\
$\chi_{ij}$ & $2$nd-order tensor perturbation in Poisson gauge & 
(\ref{metric}) \\
$\eta$ & Conformal time & (\ref{metric})  \\
$f$ & Photon distribution function & (\ref{Df}) \\
$g$ & Distribution function for massive particles & (\ref{gel}) 
\& (\ref{DG}) \\
$f^{(i)}$ & $i$-th order perturbation of the photon distribution 
function  & (\ref{expf}) \\
$f^{(i)}_{\ell m}$ & Moments of the photon distribution function 
& (\ref{angular}) \\
$C({\bf p})$ & Collision term & (\ref{collisionterm}) \& 
(\ref{Integralcolli}) \\ 
$p$ & Magnitude of photon momentum (${\bf p}=pn^i$) &  
(\ref{defp}) \\
$n^i$ & Propagation direction & (\ref{Pi}) \\
$\Delta^{(1)}(x^i,n^i,\eta)$ & First-order fractional energy 
photon fluctuations & (\ref{Delta1})  \\
$\Delta^{(2)}(x^i,n^i,\eta)$ & Second-order fractional energy 
photon fluctuations & (\ref{Delta2}) \\
$n_e$ & Electron number density & (\ref{defne}) \\
$\delta_e (\delta_b)$  & Electron (baryon) density perturbation 
& (\ref{deltac1}) \\
${\bf k}$ & Wavenumber & (\ref{BF}) \\
$v_m$ & Baryon velocity perturbation & (\ref{vzero}) \& (\ref{vdec}) \\ 
$v^{(2)i}_{\rm CDM}$ & Cold dark matter velocity & (\ref{mcc2CDM}) \\
$v^{(2)i}_\gamma$ & Second-order photon velocity & (\ref{vp2}) \\
$S_{\ell m}$ & Temperature source term & (\ref{H}) \\  
$\tau$ & Optical depth & (\ref{deftau}) \\
${\bar \rho}_\gamma ({\bar \rho}_b)$ & Background photon (baryon) 
energy density & (\ref{mcc1}) \\ 
\hline\hline
\end{tabular}
\end{table}

In these lecture notes we derived the equations which allow to evaluate 
CMB anisotropies, 
by computing the Boltzmann equations describing the evolution of the 
baryon-photon fluid up to second order. This allows to follow the 
time evolution of CMB anisotropies (up to second order) on all scales, 
from the early epoch, when the cosmological perturbations were generated,
to the present time, through the recombination era. 
The dynamics at second order is particularly important when dealing with 
the issue of non-Gaussianity in CMB anisotropies. Indeed, 
many mechanisms for the generation of the
primordial inhomogeneities  predict a level of non-Gaussianity in the 
curvature perturbation which might be detectable by present 
and future experiments. 
\\
\\
\noindent
{\bf Acknowledgments}
A.R. is on leave of absence from INFN, Sezione di Padova. S.M. 
is partially supported by INAF. 

%%%%%%%%%%%%%%%%%%%%%%%%%%%%%%%%%%%%%%%%%%%%%%%%%%%%%%%%%%%%%%%%%%%%%%%%%
%APPENDIX
%%%%%%%%%%%%%%%%%%%%%%%%%%%%%%%%%%%%%%%%%%%%%%%%%%%%%%%%%%%%%%%%%%%%%%%%%
  
\appendix
\setcounter{equation}{0}
\def\theequation{A.\arabic{equation}}

\section{Einstein's equations}
\label{LH2A}

In this Appendix we provide the necessary expressions to 
deal with the gravitational part of the problem we are 
interested in, namely the second-order CMB anisotropies generated 
at recombination as well as the acoustic oscillations of the 
baryon-photon fluid. The first part of the Appendix 
contains the expressions for the metric, connection coefficients and 
Einstein's tensor perturbed up to second order around a flat 
Friedmann-Robertson-Walker background, the energy-momentum 
tensors for massless (photons) and 
massive (baryons and cold dark matter) particles, 
and the relevant Einstein's equations. The second part deals with the 
evolution equations and the solutions for the second-order 
gravitational potentials in the Poisson gauge. 

%%%%%%%%%%%%%%%%%%%%%%%%%%%%%%%%%%%%%%%%%%%%%%
\subsection{The metric tensor}
%%%%%%%%%%%%%%%%%%%%%%%%%%%%%%%%%%%%%%%%%%%%%%
As discussed in Section 3, we write our second-order metric in the
Poisson gauge, 
\begin{equation}
\label{LH2metric}
ds^2=a^2(\eta)\left[
-e^{2\Phi} d\eta^2+2\omega_i dx^i 
d\eta+(e^{-2\Psi}\delta_{ij}+\chi_{ij}) dx^i dx^j
\right]\, ,
\end{equation}
where $a(\eta)$ is the scale factor as a function of 
conformal time $\eta$, and $\omega_i$ and $\chi_{ij}$ 
are vector and tensor peturbation modes 
respectively. 
Each metric perturbation is expanded into a 
linear (first-order) and a second-order part, as discussed in Section 3. 

%%%%%%%%%%%%%%%%%%%%%%%%%%%%%%%%%%%%%%%%%%%%%%%%%%%%%%%%%%%
\subsection{The connection coefficients}
%%%%%%%%%%%%%%%%%%%%%%%%%%%%%%%%%%%%%%%%%%%%%%%%%%%%%%%%%%
\noindent 
For the connection coefficients we find
\begin{eqnarray}
\Gamma^0_{00}&=& {\mathcal H}+\Phi'\, ,\nonumber\\
\Gamma^0_{0i}&=& \frac{\partial\Phi}{\partial x^i}+
{\mathcal H}\omega_i\, ,\nonumber\\
\Gamma^i_{00}&=& \omega^{i'}+{\mathcal H}\omega^i+e^{2\Psi+2\Phi}
\frac{\partial\Phi}{\partial x_i}
\, ,\nonumber\\
\Gamma^0_{ij}&=& -\frac{1}{2}\left(\frac{\partial \omega_j}{\partial x^i}+
\frac{\partial \omega_i}{\partial x^j}\right)+e^{-2\Psi-2\Phi}
\left({\mathcal H}-\Psi'\right)\delta_{ij}+\frac{1}{2}\chi_{ij}'+
{\mathcal H}\chi_{ij}
\, ,\nonumber\\
\Gamma^i_{0j}&=&\left({\mathcal H}-\Psi'\right)\delta_{ij}+
\frac{1}{2}\chi_{ij}'+\frac{1}{2}\left(\frac{\partial \omega_i}{\partial x^j}-
\frac{\partial \omega_j}{\partial x^i}\right)\, ,\nonumber\\
\Gamma^i_{jk}&=&-{\cal H}\omega^i\delta_{jk}-
\frac{\partial\Psi}{\partial x^k}\delta^i_{~j}-
\frac{\partial\Psi}{\partial x^j}\delta^i_{~k}+
\frac{\partial\Psi}{\partial x_i}\delta_{jk} \nonumber \\
& + & 
\frac{1}{2} \left(\frac{\partial\chi^i_{~j}}{\partial x^k}+
\frac{\partial\chi^i_{~k}}{\partial x^j} - 
\frac{\partial\chi_{jk}}{\partial x_i} \right) \, .
\end{eqnarray}

%%%%%%%%%%%%%%%%%%%%%%%%%%%%%%%%%%%%%%%%%%%%%%%%%%%%%%%%
\subsection{Einstein tensor}
%%%%%%%%%%%%%%%%%%%%%%%%%%%%%%%%%%%%%%%%%%%%%%%%%%%%%%%
\noindent
The components of Einstein's tensor read 
\begin{eqnarray}
\label{LH200}
G^0_{~0}&=&-\frac{e^{-2\Phi}}{a^2}\left[3{\mathcal H}^2-6{\mathcal H}\Psi'
+3(\Psi')^2 \right. \nonumber \\
&-& \left. e^{2\Phi+2\Psi}\left(\partial_i\Psi\partial^i\Psi-2\nabla^2
\Psi\right)\right]\, ,\\
\label{LH2i0}
G^i_{~0}&=&2\frac{e^{2\Psi}}{a^2}\left[\partial^i\Psi'+\left({\mathcal H}-
\Psi'\right)\partial^i\Phi\right]-\frac{1}{2a^2}\nabla^2\omega^i \nonumber 
\\ 
&+& \left(4{\mathcal H}^2-2\frac{a''}{a}\right)\frac{\omega^i}{a^2}\, ,\\
\label{LH2ij}
G^i_{~j}&=&\frac{1}{a^2}\biggl[e^{-2\Phi}\biggl({\mathcal H}^2-2\frac{a''}{a}-
2\Psi'\Phi'-3(\Psi')^2+2{\mathcal H}\bigl(\Phi'+2\Psi'\bigr)
\nonumber \\
&+& 2\Psi''\biggr)
+ e^{2\Psi}\biggl(\partial_k\Phi\partial^k\Phi+\nabla^2\Phi-\nabla^2\Psi
\biggr)\biggr]\delta^i_j 
+ \frac{e^{2\Psi}}{a^2} \nonumber \\
&\times& \biggl(-\partial^i\Phi\partial_j\Phi  -
\partial^i\partial_j\Phi+\partial^i\partial_j\Psi-\partial^i\Phi\partial_j\Psi
+\partial^i\Psi\partial_j\Psi-\partial^i\Psi\partial_j\Phi\biggr)\nonumber\\
&-&\frac{{\mathcal H}}{a^2}\left(\partial^i\omega_j+\partial_j\omega^i\right)
-\frac{1}{2a^2}\left(\partial^{i}\omega_j'+\partial_j\omega^{i'}\right)
\nonumber \\
&+& \frac{1}{a^2}\left(
{\mathcal H}\chi^{i'}_j+\frac{1}{2}\chi_j^{i''}-\frac{1}{2}\nabla^2
\chi^i_j\right)\, .
\end{eqnarray}
Taking the traceless part of Eq.~(\ref{LH2ij}), we find
$\Psi-\Phi={\cal Q}$, 
where ${\cal Q}$ is defined by 
$\nabla^2{\cal Q}=-P+3N$, 
with $P\equiv P{^i}_{~i}$, 
\begin{equation}
P^i_{~ j} = \partial^i\Phi\partial_j\Psi+\frac{1}{2} \left( 
\partial^i \Phi\partial_j\Phi- \partial^i \Psi\partial_j\Psi\right) 
+4\pi G_{\rm N}a^2e^{-2\Psi}T^i_{~j}
\end{equation}
and $\nabla^2 N=\partial_i\partial^j P^i_{~ j}$. 

The trace of Eq.~(\ref{LH2ij}) gives
\begin{eqnarray}
\label{LH2trace}
& & e^{-2\Phi}\left({\mathcal H}^2-2\frac{a''}{a}-2\Phi'\Psi'-3(\Psi')^2+
2{\mathcal H}\left(3\Psi'-{\cal Q}'\right)+2\Psi''\right) \nonumber \\
&&
+\frac{e^{2\Psi}}{3}\left(2\partial_k\Phi\partial^k\Phi+
\partial_k\Psi\partial^k\Psi-
2\partial_k\Phi\partial^k\Psi+2(P-3N)\right) \nonumber \\
& & = \frac{8\pi G_{\rm N}}{3} a^2 T^k_{~k}\, .
\end{eqnarray}

From Eq.~(\ref{LH2i0}), we may deduce an equation for $\omega^i$
\begin{eqnarray}
\label{LH2omegai}
&-&\frac{1}{2}\nabla^2\omega^i
+\left(4{\mathcal H}^2-2\frac{a''}{a}\right)\omega^i \\
&=&
-\left(\delta^i_j-\frac{\partial^i\partial_j}{\nabla^2}\right)
\left(2\left(\partial^j\Psi'+\left({\mathcal H}-
\Psi'\right)\partial^j\Phi\right)-8\pi G_{\rm N}a^2 e^{-2\Psi} 
T^j_{~ 0}\right)\, . \nonumber
\end{eqnarray}

\subsection{Energy-momentum tensor}
\subsubsection{Energy-momentum tensor for photons}

The energy-momentum tensor for photons is defined as

\begin{equation}
T^\mu_{\gamma ~\nu}=\frac{2}{\sqrt{-g}}\int \frac{d^3 P}{(2\pi)^3}\, 
\frac{P^\mu P_\nu}{P^0}\, f\, ,
\end{equation}
where $g$ is the determinant of the metric (\ref{LH2metric}) and $f$
is the distribution function. We thus obtain
\begin{eqnarray}
\label{LH2T00photons}
T^0_{\gamma ~0}&=&-\bar{\rho}_\gamma
\left(1+\Delta_{00}^{(1)}+\frac{\Delta_{00}^{(2)}}{2}\right)\, ,\\
\label{LH2Ti0photons}
T^i_{\gamma ~0}&=&-\frac{4}{3}e^{\Psi+\Phi}\bar{\rho}_\gamma\left(
v_\gamma^{(1)i} +
\frac{1}{2}v_\gamma^{(2)i}+\Delta^{(1)}_{00} v_\gamma^{(1)i}\right)+\frac{1}{3}
\bar{\rho}_\gamma e^{\Psi-\Phi}\omega^i \\
\label{LH2Tijphotons}
T^i_{\gamma ~j}&=& \bar{\rho}_\gamma\left(\Pi^i_{\gamma ~j}+
\frac{1}{3}\delta^i_{~j}\left(1+\Delta_{00}^{(1)}+
\frac{\Delta_{00}^{(2)}}{2}\right)
\right)
\, ,
\end{eqnarray}
where $\bar{\rho}_\gamma$ is the background energy density of photons
and 
\begin{equation}
\Pi^{ij}_{\gamma}=\int\frac{d\Omega}{4\pi}\,\left(n^i n^j-\frac{1}{3}
\delta^{ij}\right)\left(\Delta^{(1)}+\frac{\Delta^{(2)}}{2}\right)\, ,
\end{equation}
is the quadrupole moment of the photons.

\subsubsection{Energy-momentum tensor for massive particles}

The energy-momentum tensor for massive particles of mass $m$, 
number density $n$  and degrees of freedom $g_d$
\begin{equation}
T^\mu_{m ~\nu}=\frac{g_d}{\sqrt{-g}}\int \frac{d^3 Q}{(2\pi)^3}\, 
\frac{Q^\mu Q_\nu}{Q^0}\, g_m\, ,
\end{equation}
where $g_m$ is the distribution function. We obtain
\begin{eqnarray}
\label{LH2T00massive}
T^0_{m~0}&=&-\rho_m=-\bar{\rho}_m\left(1+\delta^{(1)}_m+
\frac{1}{2}\delta^{(2)}_m
\right) \;, \\
\label{LH2Ti0massive}
T^i_{m~0}&=&-e^{\Psi+\Phi}\rho_m v_m^{i}=
-e^{\Phi+\Psi}\bar{\rho}_m\left(v_m^{(1)i}+
\frac{1}{2}v_m^{(2)i}+\delta^{(1)}_m v_m^{(1)i}\right) \\
\label{LH2Tijmassive}
T^i_{m~j}&=& \rho_m\, \left(\delta^i_{~ j} 
\frac{T_m}{m}+v_m^{i} v_{m~j}\right)=
\bar{\rho}_m\left(\delta^i_{~ j} \frac{T_m}{m}+v_m^{(1)i} v^{(1)}_{m~j}\right)
\, , 
\end{eqnarray}
where $\bar{\rho}_m$ is the background energy density of massive
particles and we have included the equilibrium temperature $T_m$.

\setcounter{equation}{0}
\def\theequation{B.\arabic{equation}}

\section{First-order solutions of Einstein's equations in various eras}
\label{LH2B}

\subsection{Matter-dominated era}
\label{LH2Appmatter}

During the phase in which the CDM is dominating the energy density
of the Universe, $a\sim \eta^2$ and we may use Eq.~(\ref{LH2trace}) to
obtain an equation for the gravitational potential at first order
in perturbation theory (for which $\Phi^{(1)}=\Psi^{(1)}$)

\begin{equation}
\label{LH2PhiCDM}
\Phi^{(1)''}+3{\mathcal H}\Phi^{(1)'}=0\, ,
\end{equation}
which has two solutions $\Phi^{(1)}_+=$ constant and 
$\Phi^{(1)}_{-}={\mathcal H}/a^2$.
At the same order of perturbation theory, the CDM velocity can be read off
from Eq.~(\ref{LH2i0})

\begin{equation}
\label{LH2velocitymatter1}
v^{(1)i}=-\frac{2}{3{\mathcal H}}\partial^i\Phi^{(1)}\, .
\end{equation}

The matter density contrast $\delta^{(1)}$ satisfies the first-order
continuity equation

\begin{equation}
\delta^{(1)'}=-\frac{\partial v^{(1)i}}{\partial x^i}=-
\frac{2}{3{\mathcal H}}\nabla^2\Phi^{(1)}\, .
\end{equation}
Going to Fourier space, this implies that

\begin{equation}
\label{LH2deltamatter1}
\delta^{(1)}_k=\delta^{(1)}_k(0)+\frac{k^2\eta^2}{6}\Phi^{(1)}_k\, ,
\end{equation}
where $\delta^{(1)}_k(0)$ is the initial condition
in the matter-dominated period.

\subsection{Radiation-dominated era}

We consider a universe dominated by photons and massless neutrinos. 
The energy-momentum tensor 
for massless neutrinos has the same form as that for photons. 
During the phase in which radiation  is dominating the energy density
of the Universe, $a\sim \eta$ and 
we may combine Eqs. (\ref{LH200}) and (\ref{LH2trace}) to
obtain an equation for the gravitational potential $\Psi^{(1)}$ at first order
in perturbation theory

\begin{eqnarray}
\label{LH2firstrad}
\Psi^{(1)''}+4{\mathcal H}\Psi^{(1)'}-\frac{1}{3}\nabla^2\Psi^{(1)}
&=& {\mathcal H}Q^{(1)'}+\frac{1}{3}\nabla^2Q^{(1)}
\, ,\nonumber\\
\nabla^2Q^{(1)}&=& \frac{9}{2}{\mathcal H}^2
\frac{\partial_i\partial^j}{\nabla^2}\Pi^{(1)i}_{T~~~j}\, ,
\end{eqnarray}
where the total anisotropic stress tensor is   
\begin{equation}
\Pi^{i}_{T~j}=\frac{\bar{\rho}_\gamma}{\bar{\rho}_T}\, \Pi^{i}_{\gamma~j}
+\frac{\bar{\rho}_\nu}{\bar{\rho}_T}\, \Pi^{i}_{\nu~j}\, .
\end{equation}

We may safely neglect the quadrupole and solve Eq.~(\ref{LH2firstrad}) 
setting $u_\pm=\Phi_\pm^{(1)}\eta$. 
Then Eq.~(\ref{LH2firstrad}), in Fourier space, becomes
\begin{equation}
u''+\frac{2}{\eta}u'+\left(\frac{k^2}{3}-\frac{2}{\eta^2}\right)u=0\, .
\end{equation}
This equation has as independent solutions
$u_+=j_1(k\eta/\sqrt{3})$, the spherical Bessel function of order 1, and 
$u_{-}=n_1(k\eta/\sqrt{3})$, 
the spherical Neumann function of order 1. The latter blows up
as $\eta$ gets small and we discard it on the basis of initial 
conditions. The final solution is therefore 
\begin{equation}
\label{LH2Phir}
\Phi_k^{(1)}=3\Phi^{(1)}(0)\frac{\sin(k\eta/\sqrt{3})-
(k\eta/\sqrt{3})\cos(k\eta/\sqrt{3})}{(k\eta/\sqrt{3})^3}\, 
\end{equation}
where $\Phi^{(1)}(0)$ represents the initial condition deep in the
radiation era.

At the same order in perturbation theory, the radiation
velocity can be read off from Eq.~(\ref{LH2i0})
\begin{equation}
\label{LH20ir}
v^{(1)i}_{\gamma}=-\frac{1}{2{\mathcal H}^2}\frac{\left(a\partial^i
\Phi^{(1)}\right)'}{a}\, .
\end{equation}

%%%%%%%%%%%%%%%%%%%%%%%%%%%%%%%%%%%%%%%%%%%%%%%%%%%%%%%%%%%%%%%%%%%%%%%%

%%%%%%%%%%%%%%%%%%%%%%%%%%%%%%%%%%%%%%%%%%%%%%%%%%%%%%%%%%%%%%%%%%%%%%%%
\end{document}